\documentclass{appolb}

\usepackage{amssymb}
\usepackage{graphicx}
\usepackage{amsmath}
\usepackage{epsfig}
\usepackage{amsmath} 
\usepackage{mathtools} 
\usepackage{slashed}

\usepackage[toc,page,header]{appendix}
\usepackage{minitoc}
\usepackage{comment}
\usepackage[numbers,sectionbib,sort&compress]{natbib}

\begin{document}

\date{\today}

\title{Particle seismology: mechanical and gravitational properties from parton-hadron duality
  \thanks{Lectures given by ERA at the 65-th Cracow School of Theoretical Physics, Zakopane, Poland, 14-21 June 2025}    \thanks{Supported by MICIU (Spain) under grant No. PID2023.147072NB.I00 and Junta de Andaluc\'\i a FQM225.}}
\author{Enrique Ruiz Arriola$^{1}$\thanks{earriola@ugr.es} and Wojciech Broniowski$^{2}$\thanks{Wojciech.Broniowski@ifj.edu.pl}
  \address{$^{1}$Departamento de F\'{\i}sica At\'{o}mica, Molecular y Nuclear and Instituto Carlos I \\ de  F{\'\i}sica Te\'orica y Computacional}
\address{$^{2}$The H. Niewodnicza\'nski Institute of Nuclear Physics, Polish Academy \\ of Sciences, 31-342~Cracow, Poland}
}

\maketitle

\begin{abstract}
  The internal structure of hadrons is characterized by form factors
  which correspond to matrix elements of currents.  Among
  those, the stress-energy-momentum tensor is a universally conserved quantity providing the gravitational form factors, from which mechanical
  properties may be derived via the response to the space-time
  fluctuations.  They have received much attention because of their
  role as moments of the Generalized Parton Distributions, where the stress-energy-momentum tensor couples to
  two photons, and more recently, due to the explicit lattice QCD determination
  for the pion and nucleon. In these
  lectures we attempt a pedagogical review of the topic from a purely hadronic
  point of view, based on the notion of dispersion relations, meson
  dominance, and parton-hadron duality. We show that despite the overwhelming
  simplicity of the approach, a rather successful description of the lattice QCD data is achieved.
\end{abstract}

\newpage

\tableofcontents 

\section{Introduction}

Hadrons are extended quantum objects which feel the strong interaction
and have a variety of properties such as mass, spin, charge, radii,
magnetic moments, etc., which ultimately characterize them. The finite
extension, typically about $0.5$--$1$~fm, suggests that these properties
actually correspond to integrated extended distributions which certainly are not
homogeneous or isotropic.  As a general principle, they
become distinctly accessible by noting that hadrons behave differently
under different external perturbations. These perturbations
must be small enough such that the back reaction on the external
perturbation can be neglected, but simultaneously large enough in order to provide a
measurable cross section in a scattering process. 

Since the discovery of the internal structure of the proton by
Hofstadter~\cite{Chambers:1956zz,Hughes:1965zza}, the main source of
experimental information on intrinsic properties of hadrons has been
the electron scattering. This allows one to figure out the electric
and magnetic distributions under the assumption of the one-photon
exchange. Likewise, the neutrino and muon scattering makes it possible
to determine the axial and pseudoscalar distributions under the
one--$Z^0$ or $W^\pm$ boson exchange. On the opposite extreme,
distributions associated with strong hadronic probes, often described
by the pion exchange, are difficult to assess since they distort the
probe strongly! Finally, the gravitational interaction, characterized
by the one-graviton exchange mechanism, would provide the energy
density, pressure, and stress distributions inside the
hadron. However, the gravitational interaction, which couples to all
objects, is so small that it does not produce any measurable response,
hence a one-graviton exchange remains a gedanken process.  Thus, a
question arises: how can one determine the mechanical and
gravitational properties of hadrons, such as the mass, momentum, or
pressure densities, without ever explicitly using the
gravitons?\footnote{We disregard here the old problem of quantizing
  gravity as a fundamental theory in a consistent
  manner~\cite{Feynman:1963ax,DeWitt:1967uc}, noting that the
  effective field theory (EFT) approach is
  sufficient~\cite{Donoghue:1994dn} (see \cite{Buoninfante:2024yth}
  for a recent overview).}
  
In these lectures we concentrate on the mechanical properties of
hadrons and the corresponding gravitational form factors (GFFs),
focusing in particular on the pion and the nucleon as prominent
examples. While the GFF concept is rather old, with many notable
studies done in the 
past~\cite{Kobzarev:1962wt,sharp1963asymptotic,Pagels:1966zza,Raman:1971jg,Hare:1972pa,Truong:1989my,Gasser:1990bv,Donoghue:1991qv}, it has until recently
not been pursued  quantitatively and realistically due to the lack of experimental data for the reasons mentioned
above. 

However, the field experienced a renaissance after the proposals of
studying the deeply virtual Compton scattering (DVCS) in terms of the
Generalized Parton Distributions (GPDs), from which GFFs arise as
moments~\cite{Ji:1994av,Ji:1996nm} in the Bjorken $x$ variable. Then,
the D-term was discovered~\cite{Polyakov:1999gs} and a semiclassical
interpretation was put
forward~\cite{Polyakov:2002yz,Polyakov:2018zvc}.

A second
renaissance was triggered by the MIT group's~\cite{Hackett:2023rif, Hackett:2023nkr}  {\it
  direct} lattice QCD computation of
the relevant matrix elements in the space-like region $0
\le -t \le 2 ~{\rm GeV}^2$ with an almost physical pion mass (170~MeV), where a benchmarking $5\%$ precision for
the nucleon and the pion GFFs has been reached.
This greatly improved the seminal studies of the quark
parts~\cite{Brommel:2007zz,QCDSF:2007ifr}, recently redone with
$m_\pi = 250$~MeV~\cite{Delmar:2024vxn}, the gluonic
parts~\cite{Shanahan:2018pib}, and the gluonic trace anomaly
component~\cite{Wang:2024lrm} at larger values of $m_\pi$. 

From the phenomenological
side, a way of extracting GFFs of the pion from the $\gamma \gamma^\ast \to
\pi^0 \pi^0$ data~\cite{Belle:2015oin} was proposed
in~\cite{Kumano:2017lhr}, with further experimental prospects to emerge at Super-KEKB and ILC.
For the nucleon case, constraints on GFFs have
been obtained via DVCS from CLAS at
JLab~\cite{CLAS:2015uuo,Burkert:2018bqq}, and from the
GlueX~\cite{GlueX:2019mkq} data for the $J$/$\psi$
photoproduction~\cite{Wang:2022vhr}.   An extraction of the proton mass radius based on
photoproduction of the vector charmoniums was made in~\cite{Guo:2025jiz}.
An estimate
of the mass radius~\cite{Kharzeev:2021qkd} was computed, and a
determination of GFFs from the Compton form factors was made in~\cite{Goharipour:2025lep}.
Hadronic generalized distribution amplitudes were considered as a gateway to the time-like GFFs in~\cite{Song:2025zwl,Han:2025mvq,Han:2025eao}.
A sophisticated fit to the data was carried out in~\cite{Alharazin:2026wfh}. In~\cite{Hatta:2025ryj}, an access to GPDs from the Sullivan process was proposed.

Model calculations of the pion GFFs were carried out in numerous
approaches, including chiral effective lagrangians at small
momenta~\cite{Andrianov:1998fr,Megias:2004u,Megias:2005fj} and at
finite
momenta~\cite{Broniowski:2007si,Broniowski:2008hx,Frederico:2009fk,Masjuan:2012sk,Fanelli:2016aqc,Freese:2019bhb,Krutov:2020ewr,Xing:2022mvk,Xu:2023izo,Li:2023izn,Liu:2024jno,Liu:2024vkj,Wang:2024sqg,Sultan:2024hep,Fujii:2024rqd,Krutov:2024adh,Choi:2025rto,Puhan:2025kzz}.
    
For the nucleon, the large-$N_c$ scaling was obtained in~\cite{Goeke:2001tz}, while the leading chiral
corrections were addressed in the heavy
baryon~\cite{Belitsky:2002jp,Ando:2006sk,Diehl:2006ya,Moiseeva:2012zi} or 
covariant~\cite{Dorati:2007bk,Alharazin:2020yjv} frameworks.   Further model estimates were made in the Skyrmion~\cite{Cebulla:2007ei,Tanaka:2025pny}, 
the chiral quark soliton model~\cite{Goeke:2007fp},  the MIT bag model~\cite{Neubelt:2019sou},
the holographic QCD model~\cite{Abidin:2009hr,Mamo:2022eui}, and in AdS/QCD~\cite{Mondal:2015fok,Fujita:2022jus,Wang:2024sqg,Deng:2025fpq,Mamo:2025hur,Mamo:2026ktr}. 
The light front formulation with valence quarks was considered in~\cite{Nair:2024fit},
and a full QCD light front modeling was presented in~\cite{Xu:2024sjt}.
The QCD sum rules were applied in~\cite{Azizi:2019ytx,Anikin:2019kwi}.  A flavor decomposition within the 
light-cone sum rule approach was carried out in~\cite{Dehghan:2025ncw}. A decomposition of the nucleon 
GFFs in terms of quarks and gluons  was proposed in~\cite{Ji:2025gsq}. A chiral  soliton calculation incorporating dilaton fields was presented in~\cite{Fujii:2025aip}.
The parity doubling model was applied in~\cite{Kawaguchi:2025cuf}. A classical  model of the nucleon was investigated in~\cite{Mejia:2025oip}.
A dilaton effective theory was explored in~\cite{Stegeman:2025sca,Stegeman:2025tdl}. Finally, a dispersive determination 
was accomplished in~\cite{Cao:2024zlf}.

Importantly, the leading-order perturbative QCD (pQCD) asymptotic behavior of the pion and nucleon GFFs was 
obtained in~\cite{Tong:2021ctu,Tong:2022zax,Liu:2024vkj}.

Energy and momentum densities can be directly measured in a classical
fluid, be it gas or liquid, by placing a thermometer and barometer or
manometer inside the system. This is what is involved in the
measurement at any meteorological station used for  weather
forecasting. Clearly, this procedure is only possible for liquids and
gases. In a solid we cannot place any measuring device inside unless
we dig a hole, but one can still study the response to external
forces, namely stresses or heating. The situation in a femtoscopic
system such as a hadron is even more difficult, since the only way of
perturbing the hadron requires a space-time gravitational fluctuation
with a shorter wavelength than the hadron size.  A recent discussion
of these issues at the hadronic level is contained
in~\cite{Ji:2025qax}.

In standard Quantum Field Theory textbooks, SEM  is
routinely described in the introductory chapters as a conserved
Noether current corresponding to the symmetry of the arbitrariness of
space-time coordinates. Actually, this symmetry may be the only
continuous one characterizing the dynamics of a given system, like for instance
in the case of a neutral spin-0 particle. Because of its connection
to gravity, the topic has been regarded as a purely academic
subject. The mentioned appearance of the lattice QCD calculations provides a first
principles determination of the mechanical properties.

Mathematically, the energy and momentum are identified as group generators
of the time and space translations, whereas the angular momentum and
relativistic invariance correspond to rotations and boosts. A more
physical definition involves the inclusion of test particles, such that the
total energy and momentum are conserved if we consider the object and the
measuring device as an isolated system.

In these lectures we review in a pedagogical way some basic facts of
SEM in a variety of frameworks, from the classical
point-mechanics to the quantum field theory, with the purpose of demystify
the concept. The second part is more phenomenological and largely 
based on~\cite{Broniowski:2024oyk,RuizArriola:2024udm,Broniowski:2024mpw,Broniowski:2025ctl},
where a good deal of details and further explanations can be found.

\section{Particle seismology}
\label{sec:seism}

We first focus and review how hadron masses respond to a space-time
deformation of the constant Lorentz metric $\eta^{\mu \nu}$,
    \begin{eqnarray}
      \eta^{\mu\nu} \to     \eta^{\mu\nu} + \Delta g^{\mu\nu}(x) .
\label{eq:flat}      
    \end{eqnarray}
Clearly, the scale of this deformation must be smaller than the hadron
size. Then we can visualize it as a ``micro-earthquake'' inside the
hadron, such that the mass changes locally\footnote{Actually the
  proper terminology should probably be
  ``femto-hadron-quake''. General relativity literature often uses the
  notation for weak fields $\Delta g^{\mu\nu}(x)= h^{\mu\nu}(x)$. We
  make the distintion between the full metric $g^{\mu\nu} (x)$ and the
  constant flat metric $\eta^{\mu\nu}$ only in this section. For the
  rest of the lectures we will take $g^{\mu\nu}$ as the flat metric.}
    \begin{eqnarray}
    M \to M + \int d^4 x \Delta g^{\mu\nu}(x) \frac{\delta M}{\delta
      g^{\mu\nu}(x)}.
    \end{eqnarray}
This provides the gravitational densities and stress inside a hadron 
    \begin{eqnarray}
    T^{\mu\nu}_H (x) =  -2 \frac{\delta M}{\delta g^{\mu\nu}(x)}\Big|_{g^{\mu\nu}=\eta^{\mu\nu}} \equiv \langle H | \Theta^{\mu\nu} (x) | H \rangle .
    \end{eqnarray}
    
The physical normalized hadron state is generally described as a wave packet,
\begin{eqnarray}    
  |H \rangle = \sum_s \int d^4 p \Psi_s (p) \delta_+ (p^2-M^2) |p , s \rangle, 
\end{eqnarray}  
where  $ \delta_+(p^2-M^2) = \theta (p \cdot n) \delta(p^2-M^2) $ is
the on-shell spectral condition imposed on a given hypersurface with a normal vector $n$. 
Using the translational invariance $\Theta^{\mu\nu} (x) = e^{i P \cdot x} \Theta^{\mu\nu} (0) e^{-i P \cdot x} $ we get 
\begin{eqnarray}
T_{H}^{\mu\nu}(x) 
&=& \int d^4 p d^4 p'   e^{i x \cdot (p-p')} \delta^+(p'^2-M^2) \delta^+(p^2-M^2) \nonumber \\
&\times& \sum_{s',s} \phi_s (p')^+ 
\langle p' ,s'| \Theta^{\mu \nu} (0) |p, s \rangle \phi_s (p) ,
\end{eqnarray}
where the matrix element can be decomposed into the Lorentz irreducible and 
symmetric structures
    \begin{eqnarray}
    \langle p' , s' | \Theta^{\mu \nu} (0)| p,s \rangle = \sum_i O_i^{\mu\nu} (p',s',p,s) G_i (q^2),
    \end{eqnarray}
with $O_i^{\mu\nu} (p',s',p,s) = O_i^{\nu\mu} (p',s',p,s)$. Besides, if the total system {\it including} also
    the metric as dynamical degree of freedom does not depend on the
    particular choice of the space-time coordinates (the equivalence
    principle), then  $q_\mu O_i^{\mu\nu} (p',s',p,s)=0$. The
    Lorentz invariant coefficients $G_i(q^2) $ depend only on the
    momentum transfer due to the on-shell conditions $p^2=p'^2=M_H^2$
    and are termed as the {\it gravitational form factors} (GFFs).

    Among these form factors there is one, $D$ -- the Druck term, which turns
    out to be an intrinsic hadronic property~\cite{Polyakov:1999gs}
    associated to the conserved operator
  \begin{eqnarray}
O_D^{\mu \nu } 
  (p',p) = q^\mu q^\nu - g^{\mu \nu} q^2 \implies D(q^2), 
  \end{eqnarray}
  and which corresponds to a change of the mass against a local variation of
  the curvature, 
  \begin{eqnarray}
    D_H (x) = \frac{\delta M}{\delta R(x)} = \int \frac{d^4
      q}{(2\pi)^4} e^{i q \cdot x } D(q^2) \implies D(0)= \int d^4 x
    \frac{\delta M}{\delta R(x)},
  \end{eqnarray}
  where $R(x)= g^{\mu\nu} (x) R_{\mu\nu} (x)$ is the scalar curvature
  and $R_{\mu\nu}(x)$ the Ricci curvature tensor (see
  e.g. Ref.~\cite{Megias:2004u} for conventions) and $D(q^2)$ is the
  D-form factor. Note that close to the flat limit, Eq.~(\ref{eq:flat}), one
  has $\Delta R= (\partial^\mu \partial^\nu - \eta^{\mu\nu}
  \partial^2) \Delta g_{\mu\nu}$.  Like any form factor, this function
  is analytic in the complex $q^2$-plane except for a branch cut along
  the positive real axis, $s_0 < s < \infty$, where $s_0$ is a
  threshold (for the pion and nucleon $s_0=4m_\pi^2$).\footnote{In the
    nucleon case, there is an additional subthreshold logarithmic
    singularity at $s_a = 4 m_\pi^2 - m_\pi^2/M_N^2$, which stems from
    the $g \to \pi \pi \to N \bar N $ triangle process, and distorts
    greatly the threshold behavior of the form factor.}
    
  In QCD, $D(q^2)$ falls off faster\footnote{For mesons $D \sim 1/(q^2 \log q^2)$, whereas for baryons $D \sim 1/(q^2 \log
    q^2)^2$.}  than $1/q^2$~\cite{Tong:2021ctu,Tong:2022zax,Liu:2024vkj}, hence satisfies an unsubtracted dispersion relation
  \begin{eqnarray}
    D(t) = \frac1{\pi} \int_{s_0}^\infty ds \frac{{\rm Im} D(s)}{s-t} \implies
    D(0) = \frac1{\pi} \int_{s_0}^\infty ds \frac{{\rm Im} D(s)}{s} .
  \end{eqnarray}
 The spectral function ${\rm Im}\, D(s)$ corresponds to a
  virtual gravitational hadron-antihadron production $g^* \to H \bar
  H$ in the scalar $0^{++}$ and tensor $2^{++}$ quantum number channels.  As we
  will see later,  in QCD it is not positive definite due to the superconvergence
  sum rules. The value $D(0)$ is a gravitational property of the
  hadron, which is dynamical and cannot be deduced from a hadronic
  symmetry, similarly to the anomalous magnetic moments or the axial
  coupling constant of the nucleon. However, it is finite and
  unambiguous.  To what extent the spectral function ${\rm Im}\, D(s)$ can be determined
  in practice from our knowledge of the meson spectrum in the
  scalar $0^{++}$ and tensor $2^{++}$ channels, will be discussed later
  on (Sect.~\ref{sec:dr-sr-md}).

\section{Stress-energy-momentum tensor primer: particles}

In order to grasp the meaning of SEM, we
start from the classical particles, both non-relativistic and
relativistic, and then proceed to field theory and the interaction of
classical particles with
fields~\cite{Weinberg:1972kfs,sudarshan1974classical,barut1980electrodynamics,huang2009introduction,soper2008classical}.
The bottomline is that {\it only} point-like interactions
satisfy the SEM conservation locally with local
densities and currents.

\subsection{Classical particles}

Classical particles are characterized as being point like. Thus the
density or concentration of a particle located at point
$x_0$ is just a simple Dirac delta function,
\begin{eqnarray}
n(x)= \delta(x-x_0) .
\end{eqnarray}
Correspondingly, for a particle with mass $m$ and charge $q$, the
corresponding mass and charge densities are given by
\begin{eqnarray}
\rho_m (x) \equiv m n(x)= m \delta(x-x_0) \, , \quad \rho_q(x) \equiv q
n(x) = q \delta(x-x_0).
\end{eqnarray}
For a particle under external (conservative) forces we have
Newton's law
  \begin{eqnarray}
  m \frac{d^2 x}{dt^2} = - \nabla V(x),
  \end{eqnarray}
  from where we explicitly find the energy conservation
  \begin{eqnarray}
    E = \frac12 \left( \frac{dx}{dt}\right)^2 + V(x) \implies \frac{dE}{dt}=0.
  \end{eqnarray}
Thus, a concentration of a collection of moving particles  fulfills 
  \begin{eqnarray}
    n(x,t) = \sum_i \delta (x-x_i (t)) \implies  \partial_t n = - \sum \frac{dx_i}{dt}\nabla \delta (x-x_i(t)),
  \end{eqnarray}
which in terms of the current or flux of particles implies the continuity equation    
  \begin{eqnarray}
  \vec j(x,t) \equiv  \sum_i \frac{dx_i}{dt} \delta (x-x_i(t))   \implies
  \partial_t n + \nabla \cdot \vec j =0 .
  \end{eqnarray}  
  From here, one defines the momentum density
\begin{eqnarray}
  \vec {\cal  P} (x,t) \equiv  m \vec  j(x,t) = \sum_i m v_i \delta(x-x_i(t)),
\end{eqnarray}
  such that   
  \begin{eqnarray}
\partial_t  \vec {\cal P} (x,t) &=& \sum_i m \frac{d^2 \vec x_i}{dt^2} \delta (x-x_i(t))
    - \sum_i \frac{d \vec x_i}{dt} \frac{d \vec x_i}{dt} \cdot\nabla \delta (x-x_i(t)) \nonumber \\ 
    &=& - \vec \nabla V(x) n(x,t) - m \vec \nabla \overset{\leftrightarrow}{T} (x,t),
  \end{eqnarray}
where we have introduced the stress tensor 
  \begin{eqnarray}
  T_{ab} (x,t) = \sum_{i} \frac{d x_i^a}{dt}  \frac{d  x_i^b}{dt} \delta (x-x_i(t))
  \end{eqnarray}
  and the dyadic product notation $ \vec A
  \overset{\leftrightarrow}{T} $.  Finally, the energy density is
  naturally defined as
  \begin{eqnarray}
    {\cal H}(x,t) = \sum_i \frac12 m \left(\frac{d \vec x_i}{dt} \right)^2 \delta (x-x_i(t)) + V(x) n(x,t) , 
  \end{eqnarray}
which fulfills
 \begin{eqnarray}
\partial_t   {\cal H}(x,t) + \nabla \cdot J_E (x,t)   = - V(x) \nabla j(x,t)  
  \end{eqnarray}
in terms of the energy flux 
  \begin{eqnarray}
  \vec J_E =  \sum_i \sum_i \frac12 m \left(\frac{d \vec x_i}{dt} \right)^2
  \frac{d \vec x_i}{dt}  \delta(x-x_i(t)).
  \end{eqnarray}
  
\subsection{Phase-space point of view}

  While we are mainly interested in local quantities, formulas get
  simpler with the phase-space Hamiltonian dynamics, where
    \begin{eqnarray}
    H(p,x) = E(p) + V(x) \implies \begin{cases}
      \dot{\vec x}= \nabla_p H = \nabla_p E  \equiv \vec v \\  \\ \dot{\vec p}= - \nabla_x H = -\nabla V (x) \end{cases}.
    \end{eqnarray}
The phase-space density reads 
    \begin{eqnarray}
      W(x,p,t) = \sum_i \delta (x-x_i(t)) \delta (p-p_i(t)),
        \end{eqnarray}
which fulfills Liouville's equation 
       \begin{eqnarray}   
    \partial_t W + \partial_p H \partial_x W - \nabla_x H \partial_p W =0   
    \end{eqnarray}
and the Poisson bracket formula
    \begin{eqnarray}
     \left\{ A, B \right\} \equiv \partial_x A \partial_p B - \partial_p A \partial_x B \implies \partial_t W +  \left\{ H, W \right\}=0 .
    \end{eqnarray}
The local quantities are obtained by 
    \begin{eqnarray}   A(x,t) = \int dp A (x,p) W(x,p,t). \end{eqnarray}    
The correspondence is summarized in Table~\ref{tab:phase}. 

\begin{table}
 \begin{center}
\begin{tabular}{ |c|c|c|c|c|c|} 
 \hline
A(x,p)  & 1     &        p & H(x,p)        & $p_i p_i$ & $p_i H $\\ 
O(x,t) & n(x,t) & ${\cal P}(x,t)$ & ${\cal H}(x,t)$ & $T_{ij} (x,t)$ & $J_E(x,t)$ \\  
 \hline
\end{tabular}
\caption{Phase-space functions and the corresponding local quantities.}
\label{tab:phase}
  \end{center}
\end{table}
 
 \subsection{Relativistic particles}
 
 The previous results make the transition to the relativistic 
 Hamiltonian dynamics straightforward:
    \begin{eqnarray}
    H(p,x) = \sqrt{p^2+m^2} + V(x) \implies \begin{cases}
      \dot{\vec x}= \nabla_p H = \frac{\vec p}{\sqrt{p^2+m^2}} \equiv \vec v \\  \\ \dot{\vec p}= - \nabla_x H = -\nabla V (x) \end{cases} .   
    \end{eqnarray}
    The energy and momentum densities become
      \begin{eqnarray}
        {\cal H}(x,t) &=& \sum_i \sqrt{p_i^2+m^2} \delta(x-x_i) + n(x,t) V(x), \\ 
        {\cal P}(x,t) &=& \sum_i p_i \delta(x-x_i),
      \end{eqnarray}
with the continuity equation
      \begin{eqnarray}
           \partial_t {\cal H}(x,t) + \nabla \cdot \vec {\cal P} (x,t)  = V(x) \partial_t n(x,t)   = - V \nabla \cdot J .
     \end{eqnarray}
Relativistically,  the momentum density and the energy flux
coincide, since $\vec v = \partial_p E $ and thus $\vec p= \vec v E $~\footnote{This is not the case non-relativistically if we ignore the rest mass, i.e. we only consider the {\it kinetic} energy flux  $\vec v p^2/2m$.}.  The stress tensor is now defined as 
     \begin{eqnarray} &&T^{ab} (x,t) = \sum_i \frac{p_i^a p_i^b}{\sqrt{p_i^2+m^2}} \delta(x-x_i(t)) \nonumber \\ && \implies  
     \partial_t \vec {\cal P}(x,t) + \nabla T(x,t) = - \nabla V (x) n(x,t) \qquad .
     \end{eqnarray}
     We can combine the previous definitions into a four dimensional SEM:
    \begin{eqnarray}
    T^{\mu \nu}    = \left(\begin{matrix} {\cal H} &  \vec {\cal P} \cr
    \vec {\cal P}  & T^{ab}  \end{matrix}\right) 
    \implies \partial_\mu T^{\mu \nu} =f^\nu  ,
    \end{eqnarray}
    which is symmetric,
    \begin{eqnarray}
      T^{\mu \nu} = T^{\nu \mu}.
    \end{eqnarray}  
    Note that the trace of SEM is
given by     
\begin{eqnarray}
  T^\mu_\mu (x) =   \sum_i \frac{m^2}{\sqrt{p_i^2+m^2}} \delta(x-x_i(t)) = \epsilon- 3 p \ge 0 ,
\end{eqnarray}
where $\epsilon$ is the energy density and $p$ is the pressure,
and is manifestly positive for massive particles and zero for massless
particles.\footnote{Interestingly, this positivity condition does not in general hold in QCD. In particular, 
the pion violates it at sufficiently small distances (see Ref.~\cite{Broniowski:2024mpw}).}

\subsection{Particle interactions: non-locality and the no-go theorem}
\label{sec_no-go}

The above discussion concerned non-interacting particles in
external potentials, hence the next natural step would be to include
interactions. This is, however, not so
straightforward. In the non-relativistic case for a finite range
two-particle interaction characterized by a potential $v_{12}=
V(|\vec x_1 - \vec x_2 |)$, the local conservation law for SEM holds for
the point-like particles, but the corresponding currents are non-local {\it
  unless} the interaction has zero range~\cite{soper2008classical}. Besides, it turns out that
it is impossible to construct a Hamiltonian or Lagrangian description
of a system of interacting particles that is both relativistically
invariant and contains non-trivial
interactions~\cite{leutwyler1965no}. Thus, the only logical way out is to consider the 
external fields as dynamical~\cite{sudarshan1974classical}.

\section{Energy momentum tensor primer: fields and test particles}

\subsection{Electrodynamics and the field energy}
\label{sec:elecd}

The best (and first) known example of dynamical fields coupled to
particles is provided by classical electrodynamics and the discovery
by Poynting in 1884~\cite{10.1098/rstl.1884.0016}, where he first
realized that fields carry the energy and momentum which can be exchanged
with particles. Moreover, he also noticed that both the momentum
density and the energy flux coincide (the Poynting vector), in harmony with the
relativistic invariance. The main idea is as follows: the
particle dynamics is governed by the Lorentz force,
\begin{eqnarray}
\dot{\vec p}_i = q_i \left[ \vec E + \vec v_i \wedge \vec B \right] \, , \quad \vec p_i = \frac{m_i \vec v_i }{\sqrt{1-v_i^2}}.
\end{eqnarray}
The electric charge density and currents are now given by 
\begin{eqnarray}  
  \rho_q (x,t)&=& \sum_i q_i \delta(x - x_i(t)), \\
\vec J_q (x,t) &=& \sum_i q_i v_i \delta(x - x_i(t)),
\end{eqnarray}
and satisfy the charge continuity equation
\begin{eqnarray}
 \partial_t \rho_q (x,t) + \nabla \cdot \vec J_q (x,t) = 0.
\end{eqnarray}
The dynamical electric and magnetic fields obey Maxwell's equations (we use natural units)  
\begin{eqnarray}
 \nabla \wedge E  = - \partial_t B \, , \quad  
  \nabla \wedge B  =  J_q + \partial_t E  \, , \quad 
  \nabla \cdot E  = \rho_q \, , \quad 
  \nabla \cdot B = 0.
\end{eqnarray}
From Maxwell's equations we have, after some straightforward manipulations,
  the local conservation laws
  \begin{eqnarray}
    \partial_t {\cal H}+ \nabla \cdot {\cal P} &=& 0  \, , \\
    \partial_t {\cal P} + \nabla {\cal T} &=& 0,
  \end{eqnarray}
 where  
  \begin{eqnarray}
    {\cal H}  &=& \frac12 \left[ E^2 + B^2 \right]+ \sum_i \sqrt{p_i^2+m_i^2} \delta(x-x_i),
    \\
    \vec {\cal P} &=&   E \wedge  B + \sum_i \vec p_i \delta(x-x_i), \nonumber \\\ 
    {\cal T}_{ab} &=& E_a E_b - \frac12 \delta_{ab} E^2 + B_a B_b - \frac12 \delta_{ab} B^2 + \sum_i \frac{p_a p_b}{\sqrt{p^2+m^2}} \delta(x-x_i). \nonumber
  \end{eqnarray}
  Thus, when we place
  matter we obtain the total energy and momentum conservation,
  \begin{eqnarray}  
    \frac{d}{dt} \int d^3 x {\cal H}_{\rm field} + \sum_i v_i F_i  &=& 0  \implies {\cal H} \, \quad {\rm energy \, density}, \nonumber \\
    \frac{d}{dt} \int d^3 x {\cal P}_{\rm field} + \sum_i p_i  &=& 0  \implies {\cal P}\quad {\rm momentum \, density}. \nonumber
  \end{eqnarray}
   This allows one to identify ${\cal H}_{\rm field}$ and ${\cal P}_{\rm
     field}$ physically {\it in the vacuum} as the energy and momentum
   densities {\it without any ambiguity}. The local character of the
   SEM conservation law makes it possible to compute scattering of EM
   waves (such as the Thomson scattering). Besides, the fact that we use test particles, which due to
   relativity provides a symmetric SEM, enforces the same symmetry feature on the field piece. It is remarkable that even though particles
   interact through fields, both the total energies and momenta are {\it
     additive}. This is, however, a very special feature of electrodynamics.
   Moreover, we can identify the field energy and momentum by placing
   {\it test particles} and checking for the energy and momentum
   conservation.\footnote{This does not prevent from eventually running
     into contradictions when Maxwell's and Lorentz's equations are solved
   self-consistently; as a general rule they can only be used to
   first order perturbation theory (see, e.g., Ref.~\cite{jackson2012classical}.}

\subsection{Schr\"odinger field }

The identification of energy and momentum in Quantum Mechanics usually
comes from the correspondence principle, i.e., the fact that for
$\hbar \to 0$ Quantum Mechanics should become classical.  Here we show
that classical test particles can be added to a Schr\"odinger field in
such a way that the identification arises from the total energy and
momentum conservation.\footnote{The interaction in quantum and
  classical systems has been an object of repeated studies in the
  past, where impediments to a canonical structure have been spelled
  out~\cite{Caro:1998us}. Our setup corresponds to the interaction of
  a classical field with a classical particle.}
  
The time dependent Schr\"odinger equation in a time
independent potential $V(x)$ has two constants of motion: the probability
and the energy,
    \begin{eqnarray}
    i \partial_t \psi = - \frac1{2m}\nabla^2 \psi + V \psi  \implies
    \begin{cases}
      \frac{d}{dt} \int d^3 x |\psi|^2 =0, \\ 
      \frac{d}{dt} \int d^3 x \left[ \frac1{2m}| \nabla \psi|^2 +V(x) |\psi|^2 \right]  =0
     \end{cases} \!\!\! ,
    \end{eqnarray}
    as can be explicitly checked. The differential form of these conservation laws
can be written in terms of the   
probability density and the probability flux,
    \begin{eqnarray}
    \begin{cases}
      n(x,t) = |\psi(x,t)|^2 \\ 
      \vec J(x,t) = \frac{1}{2mi }\left[ \psi^* \nabla \psi - \nabla \psi^* \psi \right]
     \end{cases} 
      \implies  \partial_t n + \nabla \cdot \vec J =0,
    \end{eqnarray}
and the energy density and the energy flux,
    \begin{eqnarray}
  \begin{cases}  
    {\cal H}(x,t) = \frac1{2m}| \nabla \psi|^2+
    V(x) |\psi|^2 + \frac1{8m} \nabla^2 |\psi|^2 \\
    J_E (x,t) = \frac1{2m i } \left[ \nabla \psi^* \nabla^2 \psi  -  \nabla^2 \psi^* \nabla \psi \right]
 \end{cases} .   \end{eqnarray}
Considering a collection of classical particles yields also the preservation of the momentum,
  \begin{eqnarray}
&& \begin{cases}  i \partial_t \psi = - \frac1{2m}\nabla^2 \psi + \sum_i V(x-x_i) \psi \\  
  \dot{\vec p}_i = - \nabla \int d^3 x V(x-x_i(t)) |\psi(x,t)]^2
\end{cases} \\  \implies &&    
\begin{cases}
    P= \sum_i p_i + \int d^3 x m \vec J(x,t) \\ 
    E = \sum_i \frac{p_i^2}{2M} + \int d^3 x \left[ \frac1{2m}| \nabla \psi|^2+
      \sum_i V(x-x_i) |\psi|^2 \right]
    \end{cases}.
\end{eqnarray}
Note, however, that unlike the EM case, the total conserved energy is
non-additive. Finally, we note that a {\it local} SEM conservation is
not possible unless the interaction $V(x-x_i)$ has exactly zero range,
in harmony with the classical result (see Sect.~\ref{sec_no-go}). 

\subsection{Neutral Klein-Gordon field}

The last example presents an interesting case of a scalar
neutral field, where the probability is certainly not conserved, but
the energy and momentum are. The free Klein-Gordon (KG) equation reads
    \begin{eqnarray}
    (    \partial_t^2 - \nabla^2 + m^2  ) \phi=0 .
    \end{eqnarray}
Its solutions fulfill the continuity equations 
    \begin{eqnarray}
      \partial_t {\cal H}_\phi + \nabla \vec {\cal P}_\phi &=& 0, \nonumber \\
      \partial_t {\cal P}_\phi + \nabla \overleftrightarrow{\cal T}_\phi &=&0 ,  
    \end{eqnarray}    
where 
\begin{eqnarray}
      {\cal H}_\phi &=& \frac12 (\partial_t \phi)^2 + \frac12 (\nabla \phi)^2 
      + \frac12 m^2 \phi^2 \, , \nonumber \\  \vec {\cal P}_\phi &=& -\partial_t \phi \nabla \phi, \nonumber \\
      {\cal T}_{ik} &=& \nabla_i \phi \nabla_k \phi + \frac1{2} \delta_{ik}
      \left[(\partial_t \phi)^2 - (\nabla \phi)^2+ m^2 \phi^2 \right] .
\end{eqnarray}
In order to properly identify ${\cal H}$ and ${\cal P} $ as the energy
and momentum densities of the scalar field, we include test particles
that can exchange energy and momentum with the field $\phi$. This can
be done by implementing particles as a source term in the KG equation
of the form\footnote{For a Lagrangian formulation see, e.g.,~\cite{kalman1961lagrangian}.}
    \begin{eqnarray}
    (    \partial_t^2 - \nabla^2 + m^2  ) \phi= g \sum_i \sqrt{1-v_i^2} \, \delta (x-x_i) ,
\end{eqnarray}
and a single particle
Hamiltonian function of the form\footnote{These are the analogue of the Lorentz force; here we have the substitution rule $M_i \to M_i + g \phi(x_i)$.} 
          \begin{eqnarray}
    H(p,x) = \sqrt{p^2 + (M + g \phi(x))^2}  \implies \begin{cases}
      \dot{\vec x}= \nabla_p H = \nabla_p E  \equiv \vec v \\  \\ \dot{\vec p}= - \nabla_x H  \end{cases}    .   
          \end{eqnarray}          
Taking $E_i \equiv H(x_i,p_i) $, we obtain 
\begin{eqnarray}
  {\cal H} &=& {\cal H}_\phi + \sum_i E_i \delta(x-x_i), \\ 
  {\cal P}  &=& {\cal P}_\phi + \sum_i \vec p_i \delta(x-x_i), \\
  {\cal T} &=& {\cal T}_\phi + \sum_i \frac{\overleftrightarrow{p_i p_i}}{E_i} \delta(x-x_i),
\end{eqnarray}
which fulfill the continuity equation. Note that due to the local mass shift
$M_i \to M_i + g \phi(x_i)$ of the test particles, the effect is {\it
  not} additive. We get the total energy and momentum
conservation,
      \begin{eqnarray}
        E_{\rm tot} &=& \int d^3 x {\cal H}_\phi + \sum_i E_i \implies \frac{ d E_{\rm tot}}{dt}=0, \\ 
        \vec P_{\rm tot} &=& \int d^3 x {\cal P}_\phi +\sum_i  \vec p_i \implies \frac{ d \vec P_{\rm tot}}{dt}=0 ,
      \end{eqnarray}
      which entitles us to interpret ${\cal H}_\phi $, ${\cal P}_\phi$
      and as the field energy and momentum densities, whereas $ {\cal
        T}_\phi $ is the field stress {\it uniquely}. This mechanical
      balance offers a possible way of how relativistic point-like
      classical particles interact via a scalar field.

\section{Unitarity and energy vs probability conservation}

Unitarity in a scattering process is traditionally and popularly
linked to the probability conservation. Among the many interesting
properties of SEM, in this section we show that unitarity also
follows from the energy conservation. In order to stress this feature, we
consider the simplest case of the elastic scattering of a wave on a
static heavy particle target in the cases of the 
Schr\"odinger and neutral scalar fields discussed 
previously.\footnote{The EM proceeds along similar lines but becomes a bit messier due to the vector character of $\vec E$ and $\vec B$.}

\subsection{Scattering of matter waves}

The simplest relevant case appears in almost any textbook on Quantum
Mechanics and is based on the probability conservation. We discuss it here
for completeness in a particularly suitable fashion for our purposes.
Our starting point is to take a time dependent wave-packet,
    \begin{eqnarray}
    \psi(x,t ) =  \int \frac{d E}{\sqrt{2\pi}} \psi_E (x) e^{-i E t}  \implies  
    ( - \nabla^2   + U) \psi_E=  E  \psi_E .
    \end{eqnarray}
The  stationary scattering solutions become asymptotically 
    \begin{eqnarray}
      \psi_E (x) \to Z_E \left[e^{i k \cdot x} + \frac{e^{ikr}}{r} f
        \right] \equiv \psi_E^{\rm in} (x) + \psi_E^{\rm out} (x)\, , \qquad r \to \infty \label{eq:fasy} 
        \end{eqnarray}
where $Z_E$ is a suitable normalization factor and $f_E ( \hat k', \hat k)$ is the scattering amplitude for the transition, 
implied by the change of direction between the initial velocity and the observation direction $\hat k \to \hat k' \equiv \hat x$. Probability conservation implies
    \begin{eqnarray}
      \Delta N &\equiv&   N(\infty)-N(-\infty) =  \int_{-\infty}^\infty dt \frac{dN}{dt}= \int_{-\infty}^\infty dt d^3 x \partial_t \rho  \nonumber \\ &=&- \int_{-\infty}^\infty dt  d^3 x   \vec \nabla \cdot \vec {\cal J} =
  - \oint d \vec S \cdot \int_{-\infty}^\infty dt  \vec {\cal J}   =0 \,,
    \end{eqnarray}
    where  the divergence integral theorem has been used. 
    Now, using the Plancherel formula for the Fourier transformation and $d \vec S \cdot \hat r = r^2 d \Omega $, we get 
\begin{eqnarray}
     \Delta N  &=&
 \frac{i}{2m} \int dE \oint d \vec S \cdot (\psi_E^* \nabla \psi_E - \nabla \psi_E^*  \psi_E) \nonumber \\ &=&
     \frac{i}{2m} \int dE   \lim_{r \to \infty } r^2 \int d \Omega  (\psi_E^* \partial_r \psi_E - \partial_r \psi_E^*  \psi_E),
    \end{eqnarray}
   where only the asymptotic wave function enters. With the asymptotic expression~(\ref{eq:fasy}),
the limit $r \to \infty$ selects
the forward amplitude, and with averaging over the outgoing directions
   we obtain, after some manipulations,
\begin{eqnarray}
0= \Delta N=  \int d E  |Z_E|^2 \left[ -\frac{4\pi}{k}{\rm Im} f_E (\hat k, \hat k)   + \int d \Omega |f_E (\hat k, \hat x)|^2   \right]  =0 \, . \label{eq:Ncons}
\end{eqnarray}
On the other hand,
\begin{eqnarray}
  \Delta N_{\rm in} &=&   \int dE |Z_E|^2 \int d S \frac{k}m,
   \nonumber \\ \Delta N_{\rm out} &=&
     \frac{1}{m} \int dE   |Z_E|^2 \int d \Omega |f_E (\hat k, \hat x)|^2  .
  \end{eqnarray}
Taking the cross section as a probability transfer, we arrive at
    \begin{eqnarray}
    \frac{d \sigma_N}{d \Omega} = \frac{\Delta N_{\rm out}/\Delta \Omega}{\Delta N_{\rm in}/\Delta S} \implies \langle  \sigma_N \rangle = 
    \langle \frac{4\pi}{k}{\rm Im f}(\hat k, \hat k) \rangle ,
    \end{eqnarray}
where the average refers to the wave packet energy decomposition.
This is the standard well-known optical theorem result for
counting quantum particles hitting a detector.

An analogous result from the energy conservation is
    \begin{eqnarray}
      \Delta E &\equiv&   E(\infty)-E(-\infty) =  \int_{-\infty}^\infty dt \frac{dE}{dt}= \int_{-\infty}^\infty dt d^3 x \partial_t {\cal H}  \nonumber \\ &=&- \int_{-\infty}^\infty dt  d^3 x   \vec \nabla \cdot \vec {\cal P} =
  - \oint d \vec S \cdot \int_{-\infty}^\infty dt  \vec {\cal P}  =0 \,,
    \end{eqnarray}
therefore the final outcome looks very similar to Eq.~(\ref{eq:Ncons}),
with the modification that we have an additional energy factor $E$ from the energy flux expression,
\begin{eqnarray}
0= \Delta E= \int d E E |Z|^2 \left[ -\frac{4\pi}{k}{\rm Im f}(\hat k,
  \hat k) + \int d \Omega |f(\hat k, \hat x)|^2 \right]=0.
\end{eqnarray}
The relevant cross section appears now via the energy transfer (not the probability transfer)
    \begin{eqnarray}
    \frac{d \sigma_E}{d \Omega} = \frac{\Delta E_{\rm out}/\Delta
      \Omega}{\Delta E_{\rm in}/\Delta S} \implies \langle \sigma_E \rangle =
 \langle  \frac{4\pi}{k}{\rm Im f}_E (\hat k, \hat k) E \rangle = \langle \sigma_N E \rangle .
    \end{eqnarray} 
    For monochromatic wave packets 
    $|Z_E^2| = A \delta (E-E_0)$, hence $\Delta E_{\rm out} \sim E_0    \Delta N_{\rm out}$ and $ \Delta E_{\rm in} \sim E_0     \Delta N_{\rm in} $, such that 
    \begin{eqnarray}
    \frac{d \sigma_E}{d \Omega} = \frac{\Delta E_{\rm out}/\Delta \Omega}{\Delta E_{\rm in}/\Delta S}= \frac{ \int d E E |Z_E|^2 |f(\hat k, \hat x)|^2 }{\int d E E |Z_E|^2}= \frac{\Delta N_{\rm out}/\Delta \Omega}{\Delta N_{\rm in}/\Delta S}=    \frac{d \sigma_N}{d \Omega}.
    \end{eqnarray}
    As we can see, both cross sections coincide {\it only} for a
    monochromatic pulse. However, for a given broad energy spectrum
    the question arises as to what is the physical way of
    counting an event. For instance, a  calorimeter detector is just a way of
    absorbing the energy, which in a simplified picture may be viewed
    as a simple recoiling classical test system.\footnote{Actually,
      most detectors are based on the electric charge transfer.}

\subsection{Neutral and charged Klein-Gordon particle scattering }

As already mentioned, the KG equation does not possess the probability
conservation. However, it does incorporate the energy conservation and,
additionally, the charge conservation (for the charged particle case).  From
the former and in the presence of an external field $U$ we can define
an energy norm (for $U + m^2 > 0$),
      \begin{eqnarray}
      || \phi  ||_E^2  = \int d^3 x {\cal H} = \frac12 \int d^3 x  \left[ (\partial_t \phi)^2 + (\nabla \phi)^2 
      +  ( m^2 + U) \phi^2 \right] \ge 0 ,
      \end{eqnarray}      
and, correspondingly, a conserved energy scalar product
      \begin{eqnarray}
      &&\langle \phi , \varphi  \rangle_E =
      \frac12 \int d^3 x \left[ \partial_t \phi \partial_t \varphi + \nabla \phi \nabla \varphi + ( m^2 + U) \phi \varphi \right] \nonumber \\ && \implies \frac{d}{dt} \langle \phi, \varphi \rangle_E =0.               
      \end{eqnarray}
The steps to arrive to the energy-weighted 
theorems are similar as in the previous section, with
\begin{eqnarray}  \Delta E =
  - \int_{-\infty}^\infty dt \oint d \vec S \partial_t \phi \nabla \phi .
        \end{eqnarray}
For a wave packet with the scattering boundary conditions we get  
     \begin{eqnarray}
      &&\Delta E =    - \int d E \int  d \vec S  i E \phi_E (x) \nabla \phi_E (x)^* |Z_E|^2  \nonumber \\ && \to r^2 \int d E i E \int d\Omega \phi_E (x) \partial_r \phi_E (x) ^*,  \end{eqnarray}
   such that  the  (weighted) optical theorem follows:
\begin{eqnarray}
0= \Delta E=  \int dE E |Z|^2 \left[ -\frac{4\pi}{k}{\rm Im f}(\hat k, \hat k)   + \int d \Omega |f(\hat k, \hat x)|^2   \right]  =0 .\end{eqnarray}
Thus, the cross section as the energy transfer (not the probability transfer) reads 
    \begin{eqnarray}
    \frac{d \sigma_E}{d \Omega} = \frac{\Delta E_{\rm out}/\Delta \Omega}{\Delta E_{\rm in}/\Delta S} \implies \langle \sigma_E \rangle = \langle \frac{4\pi}{k}{\rm Im f}(\hat k, \hat k) E \rangle .
    \end{eqnarray} 
This particular case shows that the optical theorem for a neutral scalar
particle has to do with the energy and {\it not} the probability
conservation.

\subsection{SEM-based unitarity}

The above discussion shows that in general the SEM conservation
underlies unitarity in a purely quantum-mechanical framework. So,
rather than being an exotic object, SEM is an ubiquitous and central
quantity. In relativistic field theory, probability is not a conserved
quantity since there is no related Noether current. In QCD, for
example, one has instead color, quark number, and the SEM
conservation. As already mentioned, the only common conserved quantity
in {\it any} field theory is SEM. A formulation embodying these issues
seems to be missing, however, the issue only becomes relevant for
non-monochromatic beams.

Thus, any conservation law provides a different interpretation of
unitarity and hence of cross sections. The distinction of different
cross sections in classical transport theory is well known, where the
conventional probability cross section plays no role. In the
Fokker-Planck approximation of the linear Boltzmann equation, for
instance, only the momentum and energy transport coefficients are
physically relevant and probability is {\it not}
transported~\cite{pitaevskii2012physical}.

\section{Field Theory and local test fields \label{sec:FT}}

\subsection{Definitions}

In field theory, the canonical SEM,
$\Theta_{\mu \nu}$, amounts to the conserved Noether current
corresponding to the symmetry under the space-time
translations~\cite{Bjorken:1965zz}. In the simplest case of a scalar
field, we have for a general transformation
\begin{eqnarray}
x^\mu \to x'^\mu = x^\mu + \epsilon^\mu(x)   \implies \phi'(x') = \phi (x) \implies
\delta \phi (x) = \epsilon^\mu \partial_\mu \phi.
\end{eqnarray}
The invariance of the Lagrangian yields 
\begin{eqnarray}
\delta {\cal L} (x) &=& \epsilon^\mu \partial_\mu {\cal L} = \frac{\partial {\cal L}}{\partial \phi} \delta \phi + \frac{\partial {\cal L}}{\partial \partial^\mu \phi} \delta \partial^\mu \phi \nonumber \\ &=&  \partial^\nu \left[\frac{\partial {\cal L}}{\partial \partial^\nu \phi} \right] \delta \phi +  \frac{\partial {\cal L}}{\partial \partial^\mu \phi} \delta \partial^\mu \phi \nonumber \\ && \implies \epsilon^\nu \partial^\mu \Theta_{\mu \nu} =0 \, , \qquad 
\end{eqnarray}
therefore in the scalar theory the {\it canonical} SEM  reads 
\begin{eqnarray}
{\cal L}= \frac12 (\partial^\mu \phi)^2 -U(\phi) \implies  
\Theta^{\mu \nu} = \partial^\mu \phi \partial^\nu \phi - g^{\mu \nu}{\cal L},
\end{eqnarray}
which turns out to be symmetric $\Theta^{\nu \mu}= \Theta^{\mu \nu}$.

The canonical or Noether SEM is {\it not} always symmetric, as for
instance when dealing with the Dirac or vector fields~\cite{Freedman:1974gs,Pokorski:1987ed}.\footnote{It is
possible to redefine SEM in such a way that it becomes
symmetric. We refer the reader to old and modern (cf.~\cite{Fukushima:2026wwc} and references therein) literature for a thorough 
  discussion and interpretation of non-symmetric SEM tensors.} 
Moreover, as usual with the Noether construction, it is not unique since one may add a conserved total-derivative term,
\begin{eqnarray}
\bar \Theta^{\mu \nu}= \Theta^{\mu \nu } + \alpha \left[ \partial^\mu
  \partial^\nu - g^{\mu\nu} \partial^2\right] \phi^2,
\end{eqnarray}
where the parameter $\alpha $ is completely arbitrary. 

Given these ambiguities, the question on how one can measure
$\Theta^{\mu\nu}$ or, equivalently, how to interpret it, becomes very
pertinent, since we would naively expect a physical measurement to
be well defined. One simple and natural way is to use the test particle
concept at a given space-time location $x$, similarly to the case of 
electrodynamics discussed in Sec.~\ref{sec:elecd}. Another natural way proposed by Hilbert is via coupling
to gravity in a curved space time, in which case the flat and constant
metric is distorted, $\eta^{\mu \nu} \to g^{\mu \nu}(x) = \eta^{\mu \nu} + \delta g^{\mu \nu}(x)$, and
\begin{eqnarray}
&& \Theta^{\mu \nu} = \frac{-2}{\sqrt{-g}} \frac{\delta S}{\delta g_{\mu \nu}} \Big|_{g^{\mu\nu}= \eta^{\mu\nu}}, \nonumber \\ &&  \eta^{\mu\nu} = {\rm diag} (1,-1,-1,-1) 
\implies \Theta^{\mu \nu} = \Theta^{\nu \mu}.
\end{eqnarray}
In both cases, only the symmetric components of the SEM become
observable. We stick to this point of view in our presentation.
For the Dirac fermions, the Hilbert construction involves tetrads, as discussed in Sec.~\ref{sec:ferm}.

Coupling to gravity complies to invariance under general
transformations, $x^\mu \to x'^\mu$, therefore one has to consider an
action $S \to \int d^4 x \sqrt{-g} {\cal L}$, where the Lorentz
invariant derivatives are replaced by the world derivatives,
$\partial^\mu \phi \partial_\mu \phi \to g^{\mu \nu} \partial_\mu
\phi\partial_\nu \phi $, in a minimal way.  The ambiguous term
in $\theta^{\mu\nu}$ discussed above can be generated in a curved space-time by a
non-minimal Lagrangian ${\cal L}= \alpha R \phi^2 $, which leaves no
trace in the flat-space limit.\footnote{This is similar to the
  non-minimal gauge invariant coupling in QED, yielding an anomalous
  magnetic moment of hadrons.} Notably, it induces a change of the
Druck term
\begin{eqnarray}
D(0) \to D(0) + \alpha \label{eq:Dalpha}.
\end{eqnarray} 
As we will see below, this
ambiguity can be fixed by analyzing the production process $g \to \phi
\phi $ at high energies and ultimately has to do with taking the field
$\phi$ as fundamental or as a composite field.\footnote{This corresponds to
  implementing the SEM improvement of Callan, Coleman and
  Jackiw~\cite{Callan:1970ze}, see the discussion in Sec.~\ref{tr:anom}. For
  hadrons in QCD this is not a fundamental problem (see also the
  recent claim on the UV divergent character of D(0) for the Higgs
  boson~\cite{Beissner:2025nmg} or the arbitrariness in soliton
  models~\cite{Fukushima:2026wwc}.}

\subsection{Lorentz properties}

Under the  Lorentz group, one has the transformation law 
    \begin{eqnarray}
x^\mu \to \Lambda^\mu_\alpha x^\alpha \implies 
  \Theta^{\mu\nu} (x) \to     \Theta'^{\mu\nu} (x') = \Lambda^\mu_\alpha \Lambda^\nu_\beta \Theta^{\alpha\beta} (x),
    \end{eqnarray}
    which is a reducible representation under the trace operation; 
the trace is a scalar 
\begin{eqnarray}
\Theta (x) \equiv \Theta^\mu_\mu (x) \to  \Theta' (x') = \Theta (x).
\end{eqnarray}
On the other hand,  the (Hilbert) SEM is conserved and symmetric,
  \begin{eqnarray}
  \Theta^{\mu \nu} = \Theta^{\nu \mu} \, , \qquad \partial_\mu
  \Theta^{\mu\nu} =0 \,  \implies \text{6 independent
    components. }\end{eqnarray}
  A naive and often considered
  decomposition into traceless and traceful pieces is not consistent
  with the conservation law,
\begin{eqnarray}
&&\Theta^{\mu\nu} = \tilde \Theta_S^{\mu \nu} + \tilde \Theta_T^{\mu \nu}   \equiv \frac14 g^{\mu\nu} \Theta + \left[ \Theta^{\mu\nu} - \frac14 g^{\mu\nu} \Theta \right] 
\nonumber \\ &&\implies
\partial_\mu \tilde \Theta_S^{\mu \nu} = \partial^\nu \Theta \neq 0 .
\end{eqnarray}
A consistent decomposition, where the two tensor components are conserved
separately and are mutually orthogonal, is given by
\begin{eqnarray} \Theta^{\mu
    \nu} = \Theta_S^{\mu \nu} + \Theta_T^{\mu \nu},
\end{eqnarray}
with 
\begin{eqnarray}
\Theta_S^{\mu \nu} = \frac13 \left[g^{\mu \nu} - \frac{\partial^\mu
    \partial^\nu }{\partial^2} \right] \Theta \implies \partial_\mu
\Theta_S^{\mu \nu} =0
\end{eqnarray}
We will analyze the lattice data using this consistent decomposition. In particular, 
the meson dominance approach discussed in later sections manifestly
displays such a consistent separation explicitly.

The separation into scalar and tensor components of the SEM tensor has
also special properties under renormalization.

\subsection{Ward-Takahashi identities}

At the quantum level, the conservation laws put strong constraints on
the time ordered products, where the appearance of derivatives
requires some careful modifications in order to comply with the
Lorentz invariance. In this regard, the standard canonical approach is
rather cumbersome and plagued with the so-called Schwinger
terms~\cite{Deser:1967zzf}. We consider instead the much more transparent path
integral approach~\cite{Suura:1973xry}, where the expectation value of
a given composite field operator is written as
\begin{eqnarray}
\langle O \rangle_S = \int D\phi O e^{i S[\phi]}.
\end{eqnarray}
In particular, the time ordered product\footnote{In general, the
  so-called $T^*$ product, relevant when derivatives appear to restore the
  Lorentz invariance, corresponds to taking the derivatives {\it
    after} the $T$-operation, $T^* [ \partial_\mu \phi (x) \phi(0) ]
  \equiv \partial_\mu T[ \phi (x) \phi(0) ]$, a procedure understood
  here. We keep the $T$ symbol for a cleaner notation.} is
\begin{eqnarray}
\langle 0| T \left[ \phi(x_1) \dots \phi (x_n) \right] |0  \rangle=  \langle \phi(x_1) \dots \phi (x_n) \rangle_S = \int D\phi \phi(x_1) \dots \phi (x_n) e^{i S[\phi]}. \nonumber \\
\end{eqnarray}
The invariance under a transformation  $\phi \to \phi + \delta \phi$ implies 
\begin{eqnarray}
\delta \langle O \rangle_S = \langle  \delta O \rangle_S + \langle i \delta S O \rangle_S =0  \implies \langle  \delta O \rangle_S = - i  \langle O \delta S  \rangle_S ,
\end{eqnarray}
which is a functional form of the Feynman-Hellmann theorem. Taking
the simplest two-point function as an observable yields
\begin{eqnarray}
 \langle 0 | T \left[ \delta \phi(x_1)  \phi (x_2) \right] | 0 \rangle 
\!+\! \langle 0 | T \left[  \phi(x_1)  \delta\phi (x_2) \right] | 0 \rangle \nonumber \!=\! -i 
\langle 0 | T \left[  \phi(x_1)  \phi (x_2) \delta S \right] | 0 \rangle \, . \nonumber \\ 
\end{eqnarray}

  For a symmetry transformation with a global group generator $\delta \phi (x) = \epsilon  A \phi(x) $, with $A$ indicating an operator, the action is 
  invariant, hence $\delta S= 0$.
  The quantum Noether construction with a {\it local} group generator
  $\epsilon (x)$  takes  $\delta \phi(x) = \epsilon (x) A \phi (x) $
and $\delta S = \int d^4 x \epsilon (x) \partial^\mu J_\mu $, which yields 
\begin{eqnarray}
&&\delta (x -x_1) \langle 0 | T \left[ A \phi(x_1)  \phi (x_2) \right] | 0 \rangle 
+ \delta (x-x_2) \langle 0 | T \left[ \phi(x_1)  A \phi (x_2) \right] | 0 \rangle   \nonumber \\ && = -i 
\langle 0 | T \left[  \phi(x_1)  \phi (x_2) \partial^\mu J_\mu(x) \right] | 0 \rangle . \label{eq:wardA}
\end{eqnarray}

\subsection{Gravitational Ward-Takahashi identity for scalars}

The early work on gravitational Ward-Takahashi identities exploited the
equivalence principle in the Schwinger
formulation~\cite{Brout:1966oea,DeWitt:1967uc,Bessler:1969py}. Using  
the change of the scalar field  under general transformation $x \to x' = x + \epsilon(x) $, we get 
\begin{eqnarray}
\phi' (x') = \phi (x) \implies \delta \phi(x) = - \epsilon^\mu \partial_\mu \phi (x) \implies \delta S= \int d^4 x \epsilon^\mu \partial^\nu \Theta_{\mu\nu} \, , 
\end{eqnarray}
such that, according to Eq.~(\ref{eq:wardA}), the gravitational Ward-Takahashi identity reads 
\begin{eqnarray}
&& \delta (x -x_1) \langle 0 | T \left[ \partial^\mu \phi(x_1)  \phi (x_2) \right] | 0 \rangle 
+ \delta (x-x_2) \langle 0 | T \left[\phi(x_1)  \partial^\mu \phi (x_2) \right] | 0 \rangle  \nonumber \\ && = -i 
\langle 0 | T \left[  \phi(x_1)  \phi (x_2) \partial_\nu \Theta^{\mu\nu}(x) \right] | 0 \rangle .
\end{eqnarray}
We introduce the scalar field propagator 
  \begin{eqnarray}
i   \langle 0 | T \left[ \phi(x_1)  \phi (x_2) \right] | 0 \rangle = \int \frac{d^4 p}{(2\pi)^4} e^{i p \cdot (x_1-x_2)} \Delta(p) 
  \end{eqnarray}
and the unamputated 3-point vertex function in the momentum space, 
  \begin{eqnarray}
  \Lambda^{\mu\nu} (p',p) = \int d^4 x_1 d^2 x_2 e^{i p' \cdot x_1} e^{-i p \cdot x_2}
  \langle 0 | T \left[  \phi(x_1)  \phi (x_2) \Theta^{\mu\nu} (0) \right] | 0 \rangle .
  \end{eqnarray}
The amputated vertex function, defined as  
  \begin{eqnarray}
  \Theta^{\mu\nu} (p',p)= D(p')^{-1} \Lambda^{\mu\nu}(p',p) D(p)^{-1}  ,
  \end{eqnarray}
satisfies the following Ward-Takahashi identity in the momentum space:
\begin{eqnarray}
q_\mu \Theta^{\mu\nu} (p+q,p)= p^\nu \Delta^{-1} (p+q) - (p^\nu + q^\nu) \Delta^{-1} (p) .
\end{eqnarray}

\subsection{Gravitational form factor for scalars}

The kinematics is chosen in terms of the variables 
    \begin{eqnarray} P^\mu = \frac12 ( p^\mu + p'^\mu) \, , \qquad q^\mu = p'^\mu - p^\mu, \label{eq:kin}
    \end{eqnarray}
where the on-shell conditions are 
    \begin{eqnarray}
    p^2 = p'^2 = m^2 \implies \begin{cases} P \cdot q = 0 \\   P^2 = 4 m^2 - q^2  \end{cases} .
    \end{eqnarray}
The SEM conservation implies the on-shell condition 
    \begin{eqnarray}
    q_\mu \Theta^{\mu\nu}(p',p)=0.
    \end{eqnarray}
The gravitational form factors for a spin-0 particle  are defined via the decomposition
    \begin{eqnarray}
    \Theta^{\mu\nu} (p',p) \equiv \langle p' | \Theta^{\mu\nu} (0) | p \rangle
    = 2 P^\mu P^\nu A(q^2) +\frac{1}{2} \left ( q^\mu q^\nu - g^{\mu \nu} q^2 \right ) D(q^2). \nonumber \\
    \end{eqnarray}
The normalization condition for $A(0)$ 
comes from the Ward-Takahashi identity in the off-shell case (cf.~\cite{Broniowski:2022iip}), when we first set the on-shell condition $p^2=m^2$, and then 
approach the limit $q\to 0$. In that case 
   \begin{eqnarray} 
 q_\mu \Theta^{\mu\nu} (p,p)|_{q\to 0} &=&  2 p^\nu p \cdot q \implies \Theta^{\mu\nu} (p,p) |_{p^2=m^2}  = 2 p^{\mu} p^{\nu}  \nonumber \\ &\implies& A(0) = 1 .
   \end{eqnarray}
   
   As mentioned in the Introduction, the value of the Druck form factor at the origin,  $D(0)$,  is not constrained by symmetries and is a fundamental 
   dynamical quantity of a hadron.

\subsection{Scale transformations and trace anomaly}
\label{tr:anom}

Because of the presence of derivatives, the SEM quantum operator is badly divergent in
the ultraviolet limit. 

The scale transformations $x \to x' = e^\epsilon x$ form an abelian
group and they generate the corresponding field transformation $
\phi(x) \to \phi'(x') = e^{-\epsilon d_\phi} \phi( e^\epsilon x ) $,
where $d_\phi=1$ is the classical dimension.  For $U(\phi) = m^2
\phi^2 / 2 + g \phi^4 /4! $, the action at the classical level
undergoes the change $\delta S = \delta S_m $ with $S_m= -\int d^4 x
m^2 \phi^2/2$, such that for $m=0$ one has the scale invariance. The
corresponding dilaton current is obtained from the Noether
construction using an infinitesimal local transformation $ \delta
\phi(x)= \epsilon \left[d_\phi + x^\mu \partial_\mu \right] \phi(x)
$. Then,
\begin{eqnarray}
   D_\mu = x_\nu \bar \Theta^{\mu\nu} \implies \partial^\mu D_\mu = \bar \Theta + m^2 \phi^2
\end{eqnarray}
in terms of $\bar \Theta^{\mu \nu}$, the so-called {\it improved} SEM
tensor, which is also symmetric and conserved, 
\begin{eqnarray}
\bar \Theta^{\mu \nu}= \Theta^{\mu \nu } - \frac16 \left[ \partial^\mu
  \partial^\nu - g^{\mu\nu} \partial^2\right] \phi^2 \implies
\partial_\mu \bar \Theta^{\mu\nu}=0 . 
\end{eqnarray} 
This object was
introduced by Callan, Coleman, and Jackiw~\cite{Callan:1970ze} to
improve the high-energy behavior.  The corresponding improved action
corresponds to adding a curvature term, $\bar S = S + \int d^4x R
\phi^2/6$, and according to Eq.~(\ref{eq:Dalpha})  amounts to a change of the
Druck form factor at the origin, $D(0) \to D(0)+ 1/6$. 

Actually, at the quantum level because of the emergence of a renormalization
scale there is a scale symmetry violation. We can take $\lambda =
\mu/\mu_0$, such that the coupling constant $g$ and the mass $m$ and
the dimension $d_\phi$ changes accordingly with the renormalization  scale and a
(quantum) trace anomaly becomes
\begin{eqnarray}
\partial^\mu D_\mu = \bar \Theta =  \frac{\beta(g)}{4!}\phi^4 + m^2 (1 + \gamma_\phi) \phi^2,
 \end{eqnarray}
with $\beta(g) \equiv \partial g /\partial \log \mu  $ denoting the beta function with the anomalous
dimension
$\gamma_\phi(g) = \partial \log m /\partial \log \mu$.

\subsection{Fermion case \label{sec:ferm}}
    
The derivation for spin 1/2 particles involves the tetrad formalism, which is straightforward but a bit more involved. In the
presence of fermions,  the symmetric SEM  is defined as a functional variation
of the action~\cite{Birrell:1982ix}
\begin{eqnarray}
  \Theta^{\mu \nu} = \frac1{2e (x)}  \left[ e^\mu_A (x) \frac{\delta S}{\delta e^\nu_A (x)}
    + e^\mu_A (x) \frac{\delta S}{\delta e^\nu_A (x)} \right],
\end{eqnarray}
where $e^\mu_A (x) $ is the vierbein (tetrad) vector, with the metric tensor satisfying
$g^{\mu\nu}(x)=e^\mu_A (x) e^\nu_B (x)\eta_{AB}$, while $\eta_{AB}$ denotes the
Minkowski flat metric tensor and $e(x) ={\rm Det} [e^\mu_A (x)]$.

The covariant representation of a fermionic matrix element can be written in the form
\begin{eqnarray}
  \langle p^\prime,s^\prime| \Theta_{\mu\nu}(0) |p,s\rangle
  = \bar u(p',s') \Gamma_{\mu\nu}  u(p,s), 
\end{eqnarray}
where $u=u(p,s)$ and $u'=u(p',s')$ are the positive energy Dirac spinors with given momenta and spin projections.
The Ward-Takahashi identity has the form~\cite{PhysRev.180.1604}
\begin{eqnarray}
  q^\mu \Gamma_{\mu \nu}(p',p) &=& p'_\nu S(p)^{-1}- p_\mu S(p')^{-1} \nonumber \\
  &+& \frac{i}{2} \left[ q^{\mu} \sigma_{\mu \nu} S(p)^{-1}- q^{\nu} S(p')^{-1} \sigma_{\mu \nu} \right]
\end{eqnarray}
where $S(p)$ is the (off-shell) fermion propagator.  

Because of the Gordon identity,
\begin{eqnarray}
2m\bar u^\prime\gamma^\alpha u=\bar u^\prime(2P^\alpha+i\sigma^{\alpha\rho}q_\rho)u,
\end{eqnarray}
one can write three equivalent decompositions:
\begin{eqnarray}
  \Gamma_{\mu\nu} &=& 
      A(t)\,\gamma_{\{\mu} P_{\nu\}} 
    + B(t)\,\frac{i\,P_{\{\mu}\sigma_{\nu\}\rho}q^\rho}{2m_N}
    + D(t)\,\frac{q_\mu q_\nu-g_{\mu\nu}q^2}{4m_N} \nonumber \\
&=& \frac{1}{m_N} \left [ A(t) P_\mu P_\nu     + J(t)\,i\,P_{\{\mu}\sigma_{\nu\}\rho}q^\rho
    + D(t)\,\frac{q_\mu q_\nu-g_{\mu\nu}q^2}{4}  \right ]  \nonumber \\
&=&    2J(t)\,\gamma_{\{\mu} P_{\nu\}}  
    - B(t)\frac{P_\mu P_\nu}{m_N}
    + D(t)\,\frac{q_\mu q_\nu-g_{\mu\nu}q^2}{4m_N}, \nonumber 
\end{eqnarray}
with the relation
\begin{eqnarray}
B(t)=2J(t)-A(t). \label{eq:J}
\end{eqnarray}
We use the conventions $\sigma_{\mu\nu}=\tfrac{i}{2}[\gamma_\mu,\gamma_\nu]$ and 
$a_{\{\mu} b_{\nu\}}=\tfrac{1}{2}(a_\mu b_\nu + a_\nu b_\mu)$.
The $A(t)$ form factor is chirally even, whereas $B(t)$ and $D(t)$ are chirally odd.  Relations following from the Ward-Takahashi identity are
  \begin{eqnarray}
   A(0)=1, \qquad J(0)= \frac12 \, . \label{eq:jisr}
  \end{eqnarray}
Then relation~(\ref{eq:J}) yields $B(0)=0$. As in the scalar case, the value of $D(0)$ is
not constrained by symmetries.

\section{Gravitational form factors in QCD}

In this section we present the definition and the known features of SEM in QCD, whose matrix elements pertain directly to the 
gravitational form factors of hadrons. 

\subsection{Stress-energy-momentum tensor}
  
Following previous works on GFFs (note, however, the discussion of a
non-symmetric SEM in~\cite{Won:2025dgc}), we use the Hilbert definition of
SEM obtained from the action coupled to gravity, as discussed earlier. In the case of QCD, the Hilbert
definition coincides with the Belinfante-Rosenfeld
prescription~\cite{Pokorski:1987ed}, yielding a symmetric SEM in the
form
  \begin{eqnarray}
  \Theta^{\mu\nu} &=& \frac{i}4 \bar \Psi \left[ \gamma^\mu  \overleftrightarrow{D}^\mu +
    \gamma^\nu  \overleftrightarrow{D}^\mu \right] \Psi 
  - G^{\mu \lambda a} G^\nu_{\lambda
    a} + \frac14 g^{\mu\nu} G^{\sigma \lambda a} G_{\sigma \lambda a} + \Theta_{\rm GF-EOM}^{\mu\nu}, \nonumber \\ \label{eq:thqcd}
\end{eqnarray}  
where  $\Psi$ is the Dirac quark field carrying flavor and color, $G^{\mu \nu a}$ represents the gluon field strength tensor
with $a$ labeling the color octet representation, 
and the term $\Theta_{\rm GF-EOM}^{\mu\nu}$ denotes some extra terms from the gauge fixing and from the use of the equations of motion.

The trace anomaly of QCD is defined as the divergence of the dilatation current $D^\mu (x)$,
\begin{eqnarray}
\partial^\mu D_\mu = \Theta^\mu_\mu \equiv  \Theta  = 
\frac{\beta(\alpha)}{2\alpha}
G^{\mu\nu a} G_{\mu\nu}^a + \sum_q m_q \left[ 1 + \gamma_m (\alpha) \right] \bar q q, 
\label{eq:anom}
\end{eqnarray} 
where $q$ is the quark field of a given flavor,  $\beta(\alpha) = \mu^2 d \alpha / d\mu^2 $ denotes the QCD beta
function at the energy scale $\mu$, $\alpha=g^2/(4\pi)$ is the running coupling constant, and
$\gamma_m(\alpha) = d \log m_q / d \log \mu^2 $ is the anomalous
dimension of the current quark mass $m_q$.~\footnote{To leading order in perturbation theory one has $ \beta(\alpha) =  - \beta_0 \alpha^2/(4\pi) + \dots $ and $ \gamma_m (\alpha) = \alpha/(4\pi) + \dots  $  
  with $\beta_0= (11 N_c - 2 N_f)/3$ so that the running strong coupling constant reads
  $\alpha(\mu^2) = (4\pi /\beta_0) /\log (\mu^2 /\Lambda^2)$.}
The form of Eqs.~(\ref{eq:thqcd},\ref{eq:anom}) suggests the decomposition of SEM into the quark and gluon parts,
\begin{eqnarray}
\Theta^{\mu\nu} =  \Theta_q^{\mu\nu} + \Theta_g^{\mu\nu}, \label{eq:thdec}
\end{eqnarray}
which has been used to break up the hadron masses into various contributions~\cite{Ji:1995sv,Ji:1996ek,Leader:2013jra,Lorce:2017xzd,Hatta:2018sqd}.
The decomposition (\ref{eq:thdec}) depends on the scale $\mu$ and the renormalization scheme. It relates to the well-known feature of the 
momentum fractions carried partons in Deep Inelastic Scattering, where the momentum sum rule can be written in the form
\begin{eqnarray}
\langle p | \Theta^{\mu\nu} | p \rangle=  2 p^\mu p^\nu \left[
  \langle x \rangle_{\rm q} + \langle x \rangle_{\rm g}   \right] \implies
\langle x \rangle_{\rm q} + \langle x \rangle_{\rm g}  = 1.
\end{eqnarray}
Recent estimates~\cite{ExtendedTwistedMass:2024kjf,Alexandrou:2020sml} for the quark contributions in the pion and  nucleon states read
\begin{eqnarray}
\langle x \rangle^{\pi}_{\rm val} \sim \langle x \rangle^{N}_{\rm val} \sim 0.6 \, \qquad {\rm at}~\mu = 2~{\rm GeV} .
\end{eqnarray}

In these lectures, we do not analyze the separate contributions of quarks and gluons to GFFs, 
which are renormalization scale and scheme dependent, whereas the sum is not, in line with the applied hadronic picture.

\subsection{Ward-Takahashi identities for composite particles}

Obviously, hadrons are composite objects.  Unlike the 
cases discussed in Sec.~\ref{sec:FT}, where the
field under discussion was the the {\it same} as the one in the
Lagrangian, hadrons (mesons and baryons) are associated with Fock-state
combinations of the quark and gluon fields. This requires identifying
an interpolating field which has good quantum numbers. For instance,
the (composite) pion field is usually taken to be 
\begin{eqnarray}
  \vec \pi (x) = Z_\pi \bar q(x) i \gamma_5 \vec \tau q(x) + \dots ,  \label{eq:compose}
\end{eqnarray}
where the dots denote {\it any} combination of fields with {\it the
  same} quantum numbers, for instance $Z_\pi '\bar q q \bar q i
\gamma_5 \vec \tau q $. Here we have written only the lowest
dimensional cases. While the renormalization constants depend on the
renormalization scale $\mu$, it is believed that for typical hadronic
renormalization scales $Z_\pi \gg Z_\pi'$, such that the term included
in (\ref{eq:compose}) yields the dominant contribution. An
interesting and relevant fact is that from the point of view of the
Ward-Takahashi identities discussed above, only the transformation
properties of the hadron field in question under the corresponding
symmetry operation matter, and not the elementary or composite nature
of the state. For that reason, the Ward-Takahashi identities assume an
identical form as those from a Lagrangian with elementary fields.

\subsection{Pion GFF}

The pion being a spin-0 hadron is the simplest case, involving two form factors, $A(t)$ and $D(t)$: 
\begin{eqnarray}
\langle \pi^a(p') | \Theta^{\mu \nu}(0) | \pi^b(p) \rangle = \delta_{ab} \left [ 2 P^\mu P^\nu A(t) +\frac{1}{2} \left ( q^\mu q^\nu - g^{\mu \nu} q^2 \right ) D(t)  \right ],
\end{eqnarray}
with $a,b$ indicating the pion isospin, and the kinematics spelled out in Eq.~(\ref{eq:kin}).
The corresponding trace form factor is the combination  
\begin{eqnarray}
\hspace{-7mm} \Theta^{\mu}_\mu \equiv \Theta(q^2) = 2 \left(m_\pi^2- \frac{q^2}4 \right) A(q^2) - \frac32 q^2 D(q^2),
\end{eqnarray}
from where it follows that
\begin{eqnarray}
\lim_{q^2 \to 0} \Theta(q^2) = 2 m_\pi^2 A(0), 
\end{eqnarray} 
with $A(0)=1$, as derived previously. 

Raman~\cite{Raman:1971jg} proposed the decomposition of $\langle \pi^a(p') | \Theta^{\mu \nu}(0) | \pi^b(p) \rangle$
in terms of conserved irreducible tensors corresponding to a
well-defined total angular momentum,  $J^{PC}=0^{++}$ (scalar) and $2^{++}$ (tensor), namely
\begin{eqnarray}
\hspace{-7mm} \Theta^{\mu \nu} =  \Theta_S^{\mu \nu}+\Theta_T^{\mu \nu} \,, \qquad 
  \begin{cases}  \Theta_S^{\mu \nu} = \frac13 \left( g^{\mu \nu}-\frac{q^\mu q^\nu}{q^2}\right) \Theta \\  \Theta_T^{\mu \nu} =  2 \left[ P^\mu P^\nu - \frac{P^2}3 \left( g^{\mu \nu} - \frac{q^\mu q^\nu}{q^2} \right) \right] A
\end{cases}  .
\end{eqnarray}
Note that both parts are separately conserved, 
$q_\mu \Theta_S^{\mu \nu}=q_\mu \Theta_T^{\mu \nu}=0$, and the tensor part is traceless, $\Theta_{T \mu}^\mu=0$.
Such a decomposition was also used in~\cite{Fujita:2022jus}.
 
Since $\Theta$ and $A$ carry the information on good $J^{PC}$ channels, from the spin decomposition point of view 
they should be regarded as the primary objects, whereas the $D$ form factor, which is the key object in the mechanical considerations, 
mixes the quantum numbers, with the explicit formula
\begin{eqnarray}
D= -\frac{2}{3t} \left [ \Theta - \left ( 2 m_\pi^2 -\tfrac{1}{2}\, t \right ) A \right] . \label{eq:Dpion}
\end{eqnarray}
A chiral theorem~\cite{Novikov:1980fa,Donoghue:1991qv} states that
\begin{eqnarray}
D_\pi (0)=-1 + {\cal O} (m_\pi^2) .
\end{eqnarray}

\subsection{Nucleon GFF}

We use the covariant normalization $\bar u u =2m_N$.  Since the
operator $\Gamma_{\mu\nu}$ in question is isosinglet, we omit the
isospin index for the nucleon (the corresponding form factors are
equal for the proton and the neutron).  The condition $J(0)=1/2$ for
the nucleon is referred to as Ji's sum rule~\cite{Ji:1996nm}).

The Raman decomposition for the nucleon case takes the form 
\begin{eqnarray}
&& \hspace{-4mm} \Gamma_S^{\mu \nu} =  \frac{1}{3} Q^{\mu \nu} \Theta(t), \label{eq:RamN} \\
&& \hspace{-4mm} \Gamma_T^{\mu \nu}=  \frac{1}{m_N }\left[ P^\mu P^\nu - 
 \frac{1}{m_N } \frac{P^2}{3}  Q^{\mu \nu} \right] A(t) + 
 \frac{1}{m_N }\left [ \,i\,P^{\{\mu}\sigma^{\nu\}\rho}q_\rho -\frac{t}{6} Q^{\mu\nu} \right ] J(t), \nonumber
\end{eqnarray}
We note that analogously to the pion case, $\Gamma_S$ and $\Gamma_T$ are separately conserved, and $\Gamma_T$ is
traceless.
The trace anomaly part expressed via the other form factors reads
\begin{eqnarray}  
\Theta(t) = \frac{1}{m_N}\left[ (m_N^2 -\frac{t}{4}) A(t) - \frac{3}{4} t D(t) + \frac{1}{2} t J(t) \right ]  
\end{eqnarray}
With Eq.~(\ref{eq:jisr}), at $q=0$ we have
\begin{eqnarray}
\Theta(0)=m_N, \qquad   D(0)= \frac{4 m_N}{3} \left[ m_N A'(0)-\Theta'(0) \right].
\end{eqnarray}

\subsection{MIT lattice data}

\begin{figure}
\begin{center}
  \includegraphics[angle=0,width=.64 \textwidth]{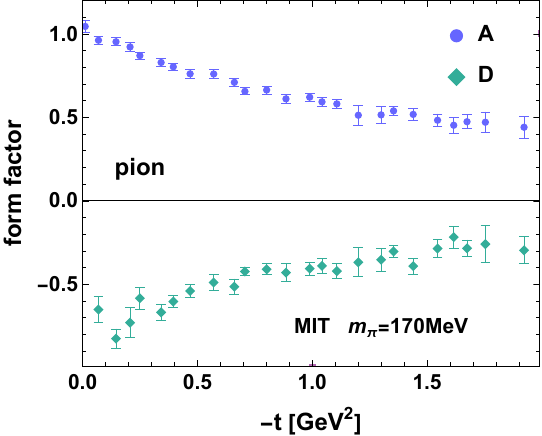} \hspace{-7mm}
  \includegraphics[angle=0,width=.36 \textwidth]{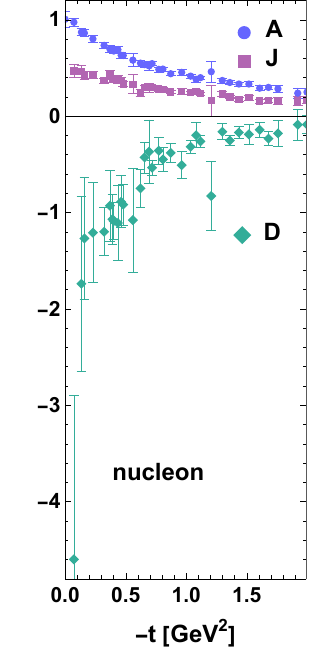}
\end{center}
 \caption{Lattice results for the gravitational form factors of the pion~\cite{Hackett:2023nkr}  (left) and nucleon~\cite{Hackett:2023rif} (right), plotted as functions of the space-like momentum transfer $-t$. \label{fig:MIT}}
\end{figure}

In recent years, much progress has been accomplished in the lattice
simulations of hadronic properties thanks to the application of the
L\"uscher-Weisz gauge action~\cite{Luscher:1984xn}. In particular, the MIT group
has been able to obtain GFFs of the pion~\cite{Hackett:2023nkr}  and the nucleon~\cite{Hackett:2023rif}  to an
unprecedented accuracy and very close to the physical point, namely
with $m_\pi=170$~MeV. The data are for the space-like momenta $0 <Q^2 <2~{\rm GeV}^2$, and obtained 
separately for the quark of three flavors and
the gluon components. As already mentioned, in these lectures, however, we use the total
(quark+gluon) quantities, since as corresponding to conserved currents, they do not depend on the renormalization
scale or scheme. As we will show, these data can be described efficiently with an ansatz whose 
parameters (meson masses) can be read off {\it directly} from the Particle Data Group tables.  
 
\section{Dispersion relations, sum rules, and meson dominance}
\label{sec:dr-sr-md}

The form factors are dynamical quantities which obey important
mathematical constraints based on analyticity in the momentum transfer
variable extended to the complex plane. For a readable introduction we
refer to~\cite{Bjorken:1965zz,barton1965introduction,k1969fields}. The bottom-line is that form
factors are described via corresponding spectral functions whose low- and high-energy behavior is theoretically known.

\subsection{Form factor and crossing}

\begin{figure}[hbt]
\begin{center}
\includegraphics[angle=0,width=.55 \textwidth]{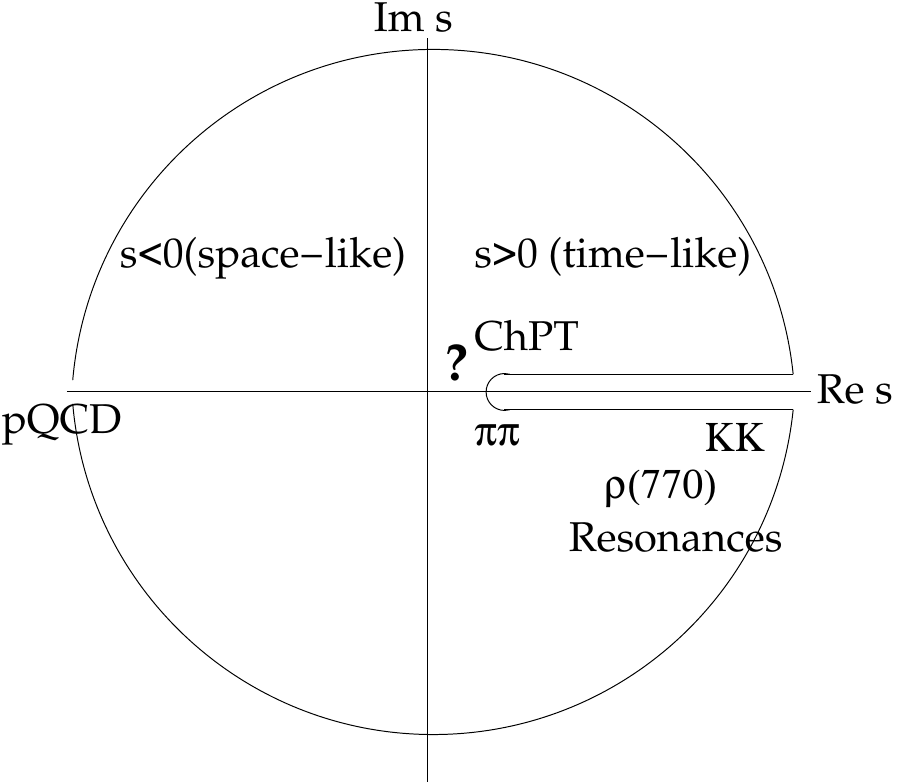}
\end{center}
\caption{Analyticity structure of the vector form factor of the pion in the complex $s=q^2$ plane. The question mark corresponds to the unphysical region $0<s<4m_\pi^2$. Labels $\pi\pi$ and $KK$ indicate the beginning of the corresponding 
two-meson production cuts (unitarity cuts). \label{fig:an}}
\end{figure}

Here, we switch for several subsections  to a much better known case of the
vector (charge) form factor of the pion, since the basic analyticity features extend analogously
to the gravitational form factors. From two specific processes, the elastic electron scattering on the pion
and the electron-positron annihilation, 
\begin{eqnarray}
&& e^- \pi^+ \to e^- \pi^+ \;  \implies \;
\langle \pi^+ (p') | J_Q^\mu (0) | \pi^+ (p) \rangle = F_\pi^Q (q^2) (p'^\mu + p^\mu)     \nonumber \\ &&  \hspace{6cm}  q^2 < 0 \, \qquad \text{space-like}, \nonumber \\
&&e^+ e^- \to \pi^+ \pi^- \;  \implies \;
 \langle \pi^+ (-p' ) \pi^- (p) |J_Q^\mu (0) |0 \rangle = F_\pi^Q (q^2)  (p'^\mu + p^\mu) \nonumber \\ &&  \hspace{6cm} q^2 > 4 m_\pi^2  \, \qquad \text{time-like},
\end{eqnarray}
one extracts the differential cross section and the total annihilation
cross section, respectively, which are related. This relation corresponds to the rotation of the
time axis in the corresponding Feynman diagram.  Analyticity of the amplitude
connects these two processes described with the {\it same
  and unique} function of the complex variable $s$ in different
domains which are experimentally accessible
(cf.~Fig.~\ref{fig:an}). This leaves the region $0 \le q^2 \le 4
m_\pi^2$ as {\it unphysical}, and it can only be reached via analytical
continuation.\footnote{We are neglecting electromagnetism here; otherwise
$\pi^+ \pi^-$ has a bound state.} The analyticity principle states that the form factor $F(s)$
can only have singularities on the real axis.

Since $F(q^2) = F(q^2)^\ast$ at
$q^2 < 0$, we infer that $ F(s^*)=F(s)^*$ (the Schwarz reflection
principle).  From there, it follows that the discontinuity is 
\begin{eqnarray}
{\rm Disc} \,F(s) = 2 i {\rm Im} F(s+ i \epsilon) \, , \qquad s > 4 m_\pi^2.
\end{eqnarray}
The function exhibits cuts starting at $s > 4 m_\pi^2$, which together
with the reflection principle guarantees the existence of only two
Riemann sheets; in the physical region they are denoted as $F_{\rm I}(s)$
and $F_{\rm II}(s)$, fulfilling $F_{\rm I}(s+i \epsilon)= F_{\rm II}(s-i \epsilon) $ 
and $F_{\rm II}(s+i \epsilon)= F_{\rm I} (s-i\epsilon) $ for $s \ge 4m_\pi^2$. Unlike $F(s) \equiv F_{\rm I}(s) $,
the analytical continuation may have singularities such as poles and
cuts. Actually, the resonances correspond to poles on the second
Riemann sheet, therefore
\begin{eqnarray}
  F_{\rm II} (s) 
  \underbrace{\to}_{s \to m_R^2 - i m_R \Gamma_R} \frac{Z_R}{s-m_R^2+i m_R \Gamma_R} + \dots.
\end{eqnarray}
The quantities $m_R$, $ \Gamma_R$, and $Z_R$ stand for the resonance
mass, width, and the residue at the pole, respectively. Thus, in the
analysis of form factors we select resonances with given quantum
numbers (in the case of the pion vector-isovector form factor, it
corresponds to the $\rho,\rho',\rho'', \dots$ states).

\subsection{Low energies}

\begin{figure}
\begin{center}
\includegraphics[angle=0,width=.6 \textwidth]{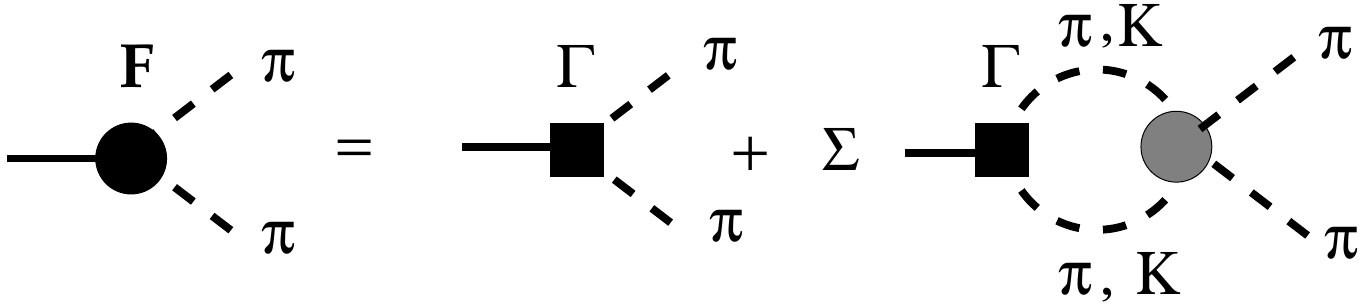} 
\end{center}
\caption{ 
Diagrammatic representation of the Bethe-Salpeter equation with the $\pi\pi$ and $KK$ production channels. \label{fig:bs}}
\end{figure}

In general terms, one has the integral equations of the Bethe-Salpeter form which in an operator form read (see, e.g.,~\cite{Nieves:1999bx}),   
\begin{eqnarray}
  F = \Gamma + \Gamma G_0 T \, , \qquad T = V + V G_0 T ,
\label{eq:bs}
\end{eqnarray}
and are supposed to hold at sufficiently low energies.
A pictorial representation in terms of {\it hadronic} Feynman diagrams is shown in Fig.~\ref{fig:bs}.
In the time-like region $e^+ e^- $ becomes {\it inelastic} for
$\sqrt{s}\ge 2m_\pi n$ ($n=1,2, \dots$) and the processes $e^+ e^- \to
\gamma^* \to n(\pi^+ \pi^-)$ occur. 

At energies below the $2\pi^+ +2\pi^-$ threshold, $s< (4 m_\pi)^2$, the 
discontinuity can be constrained from Watson's theorem (see Appendix~\ref{app:watson} for a simple proof.). In practice, all inelastic 
contributions below the $K \bar K$ threshold are neglected, such that 
\begin{eqnarray}
  \frac{F(s+ i \epsilon)}{F(s-i \epsilon)} &=&  \frac{T(s+ i \epsilon)}{T(s-i \epsilon)} \implies   F(s) = |F(s)| e^{i \delta_{11}(s)}
\end{eqnarray}
and one gets
\begin{eqnarray}
{\rm Im }  F (s)&=& |F(s)| \sin \delta_{11}(s)  > 0 , \qquad 4 m_\pi^2 < s < 4 m_K^2,
\end{eqnarray}
where the last inequality comes from the fact that $\delta_{11}(s) >0$, i.e.,  the $\pi\pi$ interaction is {\it attractive}.
From this relation, the known threshold behavior of the scattering phase shift allows one to infer the threshold behavior of the form factor itself, 
namely
  \begin{eqnarray}
 \hspace{-5mm}  \delta_{11}(s) \sim a_{11} (s/4-m_\pi^2)^\frac32  \implies {\rm Im }  F (s) \sim  |F(4m_\pi^2) | a_{11} (s/4-m_\pi^2)^\frac32.
  \end{eqnarray}
  
In the complex plane, Watson's theorem becomes 
\begin{eqnarray}
  F_{\rm II} (s) = S_{\rm II} (s) F_{\rm I} (s),  
\end{eqnarray}
where $S_{\rm II}(s)$ is the scattering matrix in the second Riemann
sheet of the complex $s$-plane. While the unitarity condition $S_{\rm II}(s) = 1/S_{\rm I} (s)$ ensures that the poles of $S_{\rm II}(s)$ are
zeros of $S_{\rm I} (s)$, this does not imply anything concerning the zeros of $F_{\rm I}(s)$.

\subsection{Large-momentum behavior from pQCD} 

Clearly, the hadronic representation is inadequate for large $Q^2$,
where a $q\bar q$ state allows for a one gluon exchange (or more
complicated processes suppressed perturbatively).  The known
leading-order (LO) pQCD asymptotic formula for the space-like vector
form factor allows one to obtain the discontinuity along the cut at
asymptotically large $s>0$,
\begin{eqnarray}
&&  F (-Q^2) \simeq \frac{16\pi F_\pi^2 \alpha_s (Q^2)}{Q^2}
  \sim \frac{64 \pi^2 F_\pi^2}{ \beta_0 Q^2 \log (Q^2/\Lambda^2)} \underbrace{\to}_{Q^2 \to e^{-i\pi} s } -\frac{64 \pi^2 F_\pi^2}{\beta_0 s ( \log(s/\Lambda^2) -i \pi) } \nonumber \\ 
&& \hspace{2cm}  \implies \frac1{\pi} {\rm Im} F(s) = - \frac{64 \pi^2 F_\pi^2}{ \beta_0 s(\log^2 (s/\Lambda^2) + \pi^2)} < 0. \label{eq:vecas}
\end{eqnarray}

\subsection{Dispersion relation}

A sketch of the complex-plane structure in the complex variable $s$ is depicted in
Fig.~\ref{fig:an}. Using Cauchy's theorem for the contour indicated in
Fig.~\ref{fig:an} and the asymptotic behavior, one obtains 
an unsubtracted dispersion relation of the form
\begin{eqnarray}
F(-Q^2)=\frac{1}{\pi} \int_{4m_\pi^2}^\infty ds \frac{{\rm Im}F(s)}{s+Q^2}, 
\end{eqnarray}
with the value at the origin normalized to the charge of $\pi^+$: 
\begin{eqnarray}
F(0)= 1 = \frac{1}{\pi} \int_{4m_\pi^2}^\infty ds \frac{{\rm Im}F(s)}{s}. 
\end{eqnarray}
Besides, a {\em superconvergent sum rule}~\cite{Donoghue:1996bt} follows form a simple observation that 
$\lim_{Q^2 \to \infty} Q^2 F(-Q^2) =0$ (cf. Eq.~(\ref{eq:vecas})). Thus
\begin{eqnarray}
\lim_{Q^2 \to \infty} Q^2 F(-Q^2) = \frac{1}{\pi} \int_{4m_\pi^2}^\infty ds {\rm Im}F(s) =0,
\end{eqnarray}
which implies that ${\rm Im}\, F(s)$ must change sign (at least once)
along the unitarity cut. Note that from Watson's theorem and the
attractive character of the $\pi\pi$ interaction in the $11$ channel,
the first zero of the spectral function must lie {\it above } the $K
\bar K$ threshold.

\subsection{Line shapes,  finite widths, and space-like momenta}

The conditions of analyticity can be solved by using specific
parameterizations which are phenomenologically motivated, particularly
in the case of resonances, whose position in the complex plane is {\it
  process independent}. However, complex energies cannot be measured,
and hence an analytical extrapolation from the experimentally
accessible real axis and the complex plane becomes mandatory.
Breit-Wigner or Gounaris-Sakurai~\cite{Gounaris:1968mw} functions are popular profiles, often used to model this energy dependence. As such, these 
models directly comply to unitarity requirements (such as
Watson's theorem).  Their applicability   can only be validated by the data, since the
separation between the resonance contribution and the background is
{\it process dependent}. Fortunately, these subtleties become rather irrelevant in
the space-like region, as we argue below.

To show this, we use the Omn\`es representation of the form factor,
\begin{eqnarray}
  F(t) = \exp \left[ \frac{t}{\pi} \int_{4m_\pi^2}^\infty \frac{ds}{s} \frac{\delta(s)}{s-t} \right], \; F(0)= 1  \implies \frac{F(t+i0)}{F(t-i0)}= e^{2 i \delta (s)},
  \label{eq:omnes}
\end{eqnarray}
which complies with Watson's theorem.\footnote{This solution is not
  unique, as we can multiply the Omn\`es function by an arbitrary
  polynomial $P(s)$. We fix it here to $P(s)=1$ for simplicity and
  discuss generalizations later.} This in principle allows one to {\it
  predict} the form factor when the elastic phase-shift is known
either experimentally or theoretically. In this context,  a resonance is
a pole in the second Riemann sheet of the scattering matrix and hence
also of the form factor,
\begin{eqnarray}
  1/S_{\rm II} (s_R) =  S_{\rm I} (s_R) =0 \implies 1/F_{\rm II} (s_R)=
  F_{\rm I} (s_R)/S_{\rm II} (s_R)=0.
\end{eqnarray}
To simplify the discussion, let us consider the energy-dependent
Breit-Wigner parametrization in the scattering region $s> 4m^2$:
\begin{eqnarray}
N(s) &=& M^2 -s + i M \Gamma f(s), \; S(s)= e^{2 i \delta (s)} = \frac{N(s)}{N(s)^*} \nonumber \\ 
\implies \delta(s)&=& \tan^{-1}\left[ \frac{M \Gamma f(s)}{M^2-s} \right] \, , \qquad  \delta(M^2)=\frac{\pi}{2} \, , \quad \Gamma= \frac{1}{M \delta'(M^2)}, \nonumber \\
\label{eq:bw-profiles}
\end{eqnarray}
where $f(M^2)=1$. In principle, the particular shape for $f(s)$
depends on the background which is process dependent.  Clearly, in the
limit of narrow width, the phase becomes $\delta(s)= \pi
\theta(s-M^2)$ and one obtains a simple monopole for the form factor,
regardless of the profile function $f(s)$,
\begin{eqnarray}
  F(t) \underbrace{\to}_{\Gamma \to 0} \exp \left[ t \int_{M^2}^\infty \frac{ds}{s} \frac1{s-t} \right]= 
\frac{M^2}{M^2-t} \implies
\frac1{\pi} {\rm Im} F(s) = \delta(s-M^2) \, .
\end{eqnarray}

\begin{figure}[hbt]
\begin{center}
   \includegraphics[angle=0,width=.49 \textwidth]{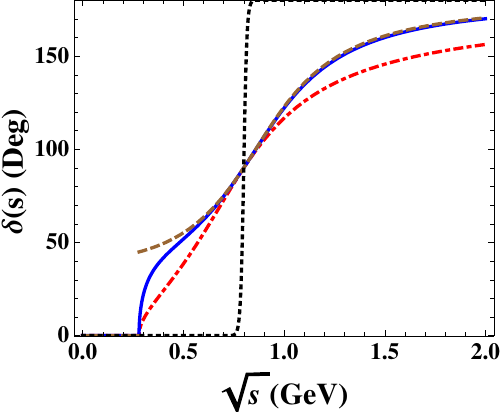}
      \includegraphics[angle=0,width=.49 \textwidth]{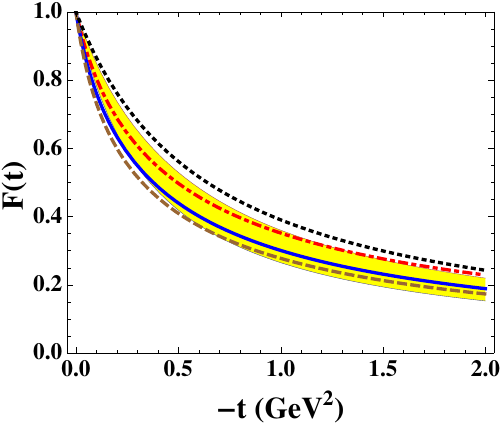}  
\end{center}
\caption{The case study of the $S$-wave phase shift $\delta(s)$, and the Omn\`es form factor in the space-like region, $F(t)$, for several 
  parameterizations (see main text):  A (brown dashed), B (blue solid), C (red dot-dashed), zero-width Monopole with $M=0.8$GeV (black dotted). The yellow band represents   monopoles with masses in the range $0.60-0.75$~GeV. 
\label{fig:BW-FF}}
\end{figure}

The question is to what extent does $F(t)$ resemble a monopole for a
{\it finite} width $\Gamma$.  We analyze this issue for a much less
favorable case: the {\it widest} known QCD resonance, namely, the
$0^{++}$ isoscalar $f_0(600)$, also called the $\sigma$-meson.  To
this end, we propose several profiles for the function in the
scattering region $s > 4 m^2$:
\begin{eqnarray}
  f_A (s) &=& 1 , \nonumber\\ 
  f_B(s)  &=& \sqrt{(s-4m^2)/(M^2-4m^2)} , \nonumber \\
  f_C(s)  &=& \sqrt{(1-4m^2/s)/(1-4m^2/M^2)}  .
\end{eqnarray}
The corresponding phase shifts are depicted in Fig.~\ref{fig:BW-FF} for the
numerical values $M=0.8$GeV and $\Gamma=0.7$GeV, and taking $m=m_\pi$.
From these profiles we may obtain, via the Omn\`es representation of
Eq.~(\ref{eq:omnes}), an analytical function in the complex
plane.\footnote{Actually, with the exception of case B, the
  corresponding functions $N(s)$ are not analytical by themselves,
  which prevents the determination of the resonance pole in the second
  Riemann sheet. This is not a problem, since the restoration of
  analyticity of the form factor by the dispersion relations can be done
  by using its phase, as we do here. } The results for the FF in the
space-like region are presented in Fig.~\ref{fig:BW-FF} for
several choices of the function $f(s)$ and with suitably chosen $M$
and $\Gamma$, which resemble qualitatively the realistic benchmark
determinations of the $\pi\pi$ scattering phase-shift form the
solution of Roy equations~\cite{Garcia-Martin:2011iqs}.

We can see that even for a broad $S$-wave resonance, and for a variety
of profiles, the form factor resembles closely a family of monopoles
for the space-like momenta. The monopole parameter is roughly the
resonance BW mass, with an uncertainty compatible with its width (it
is actually much smaller). This is a general feature which does not
depend on the chosen partial waves.  It becomes particularly helpful
when the resonance parameters are known but the phase shifts are not
so well known, as is the case away from the resonance region in many
processes. A handy possibility is provided
by the {\it halft-width rule} (see Appendix~\ref{app:hwr}).

\subsection{The large-$N_c$ limit}

It is remarkable that the large-$N_c$ limit of QCD of t'Hooft and
Witten~\cite{tHooft:1973alw,Witten:1979kh}, i.e., a limit where $N_c
\to \infty $ with $\alpha_s N_c$ fixed, provides a rationale for
the features discusses above, namely, that resonances are narrow. One can show
that
\begin{eqnarray}
\begin{cases}  M_B = {\cal O}(N_c) \, , \qquad \Gamma_B ={\cal O}(N_c^0) \\
  m_M = {\cal O}(N_c^0) \, , \qquad \Gamma_M = {\cal O}(N_c^{-1})
\end{cases} \implies \Gamma/M = {\cal O}(N_c^{-1})  .
\end{eqnarray}  
An average value of all PDG resonances~\cite{ParticleDataGroup:2024cfk} yields the so-called Suranyi's ratio~\cite{RuizArriola:2011nrw}~\footnote{The story of this ratio starts back in 1967 and it can be traced from~\cite{Masjuan:2017fzu,RuizArriola:2017ggc}.}
\begin{eqnarray}
  \left \langle \frac{\Gamma}{M} \right \rangle_{\rm exp} \equiv \sum_i g_i \frac{\Gamma_i}{M_i}= 0.12(8) ,
\end{eqnarray}
where $g_i$ is the spin, isospin and anti-particle multiplicity of a
state with resonance mass $M_i$ and width $\Gamma_i$. This ratio for
mesons and baryons numerically coincides. The uncertainty is mainly
from the corresponding variance. The numerical value corroborates the
fact that on average the QCD resonances are indeed narrow.

\subsection{The incompleteness problem}  

Measuring the pion form factor in the space-like region directly from the
electron scattering is difficult, since the charged pion decays and cannot be used as a
target. However, in the time-like region, the charged pions are created
and measured before they decay. The dispersion theory suggests that we may
obtain one process from the other by invoking analyticity.  Despite
the attractive theoretical features, purely dispersive methods
encounter unpleasant inconveniences in practice, since we only have
experimental access to modules of the form factors on a discrete and
finite grid,
\begin{eqnarray}
  |f(s_1)| , \dots , |f(s_N)| \, , \qquad 4m_\pi^2 <  s_1 < \cdots < s_N =s_{\rm max},
\end{eqnarray}  
mostly up to a limited range of momenta, be it space-like or time-like.
The phase problem may be solved using a modulus-phase decomposition,
after making an assumption on the number of zeros (see,
e.g., Ref.~\cite{RuizArriola:2024gwb}). The discrete problem is solved either by
interpolation or by using a sufficiently flexible parametrization
based on a smoothness assumption. However, the maximum upper boundary
energy problem turns out to be more serious: either pQCD applies down to
$s_{\rm max}$ or we have to give up some predictive power. 

This is illustrated by a recent dispersive
analysis of the BaBar data~\cite{BaBar:2012bdw}, for the EM pion form factor in the time-like region, where 300
points with $3$~MeV separation below $s_{\rm max}= 9~{\rm GeV}^2$ allow one to extract
the phase~\cite{RuizArriola:2024gwb} and test the fulfillment of the sum rules. From the data one gets 
\begin{eqnarray}
\frac{1}{\pi} \int_{4m_\pi^2}^{s_{\rm max}} ds
\frac{\operatorname{Im} F (s)}{s} \Big|_{\rm Data}
&=& 1.01(1)_{\textrm{st}}(^{+2}_{-1})_{\textrm{syst}},
\label{sumrule_charge_partial},  \nonumber \\
\frac{1}{\pi} \int_{4m_\pi^2}^{s_{\rm max}} ds \operatorname{Im} F (s) \Big|_{\rm Data}
&=& 0.63(2)_{\textrm{st}}(^{+7}_{-4})_{\textrm{syst}}~\textrm{GeV}^2.  
\label{eq:sumrule_asymptotic_partial}
\end{eqnarray}
As we can see, the mismatch in the charge sum rule is small, at a level of a  percent, whereas the superconvergence sum rule 
is very far from being satisfied. Here a scale for comparison is $m_\rho^2 = 0.6~{\rm GeV}^2$, hence we need a large negative value 
in Eq.~(\ref{eq:sumrule_asymptotic_partial}) from the integration 
beyond $s_{\rm max}$.

The pQCD part extrapolated down from infinity to the scale $s_{\rm max}$ yields
\begin{eqnarray}
\frac{1}{\pi} \int_{s_{\rm max}}^\infty ds
\frac{\operatorname{Im} F (s)}{s}\Big|_{\rm pQCD}
&=& - \underbrace{0.0025}_{\rm LO}- \underbrace{0.0011}_{\rm NLO}- \underbrace{0.0006}_{\rm NNLO} \, , 
\label{sumrule_charge_pqcd} \nonumber \\
\frac{1}{\pi} \int_{s_{\rm max}}^\infty ds \operatorname{Im}F (s) \Big|_{\rm pQCD}
&=& - \underbrace{0.114}_{\rm LO}- \underbrace{0.030}_{\rm NLO}- \underbrace{0.013}_{\rm NNLO} \, \textrm{GeV}^2.
\label{eq:sumrule_asymptotic_pqcd}
\end{eqnarray}
We thus note that it has a tiny negative contribution to the charge sum rule, and only about a fifth of the needed magnitude in the 
superconvergence sum rule. Therefore,  while superconvergence is a theorem, pQCD is far away, even if we use it all the way down to 
$s_{\rm max}=9~{\rm GeV}^2$.

A possible remedy is to use subtracted dispersion relations, but this requires the use of unknown subtraction constants, 
hence the predictive power is diminished. For example, with two subtractions we get explicitly
\begin{eqnarray}
F(-Q^2)= 1 - Q^2 F'(0) + \frac{1}{\pi} \left[ \int_{4m_\pi^2}^{s_{\rm max}} +
  \int_{s_{\rm max}}^\infty  
  \right] ds \frac{Q^4}{s^2}\frac{{\rm Im}F(s)}{s+Q^2},
\end{eqnarray}
where $F(0)=1$ due to the charge conservation. The last term
is ${\cal O}(Q^4/s_{\rm max}^2)$, hence it is suppressed for $Q^2 \ll
s_{\rm max}$, and we can {\it ignore} the high energy tail, but then
$F'(0)$ cannot be predicted with the data. The situation worsens for form factors where $F(0)$
is not constrained by a conservation law.

\subsection{Extended meson dominance}

At the field theoretical level, the effective narrowness of the
resonance in the space-like region can be efficiently implemented in
terms of a current-field identity proposed by Sakurai~\cite{sakurai1969currents},
\begin{eqnarray}  
  J_3^\mu  = f_\rho m_\rho^2 \rho_3^\mu. 
 \end{eqnarray}
which yields a monopole form factor for the pion 
\begin{eqnarray}
  F(-Q^2) = \frac{m_\rho^2}{m_\rho^2+Q^2} .
\end{eqnarray}
The space-like vector form factor of the pion is very well
approximated with this single meson dominance ansatz after the
half-width rule is implemented as a conservative uncertainty
estimate. This ansatz, however, does not comply strictly to the
superconvergence sum rule, an issue related to the incompleteness
problem discussed above.


At this place, after a somewhat lengthy but simpler and more pedagogical  
discussion of the pion EM form factor, we return to the properties of SEM and GFFs, the principal topic of these lectures.

The generalized meson-dominance construction for GFFs follows the derivation of the previous section, but now for different quantum number for the intermediate 
meson states. 
Saturation with the $0^{++}$ and $2^{++}$ isoscalar states (note that the Raman decomposition is manifest) yields
the structure for the currents of the form~\cite{Krolikowski:1967ryy,Raman:1970wq,Raman:1971ur} 
\begin{eqnarray}
\Theta^{\mu\nu} = \sum_S \frac13 f_S \left( \partial^\mu \partial^\nu - g^{\mu \nu} \partial^2 \right) S +
\sum_T f_T  m_T^2 T^{\mu\nu},  \nonumber 
\end{eqnarray}
where $S$ and $T^{\mu\nu}$ are scalar $0^{++}$ and tensor $2^{++}$
fields, respectively. Certainly, $T^{\mu\nu} = T^{\nu\mu}$ and $T^\mu_\mu=0$.
On-shell, they have masses $m_S $ and $m_T$, respectively, and
$\partial^\mu T_{\mu\nu}=0$ (for the complete Lagrangian see, e.g.,
\cite{Toublan:1995bk,Ecker:2007us}). Denoting the corresponding sources as $J_S$ and $J_T^{\mu \nu}$, and
using the equations of motion 
\begin{eqnarray}
(\partial^2 + m_S^2) S = J_S, \qquad (\partial^2 + m_T^2) T^{\mu\nu} = \tilde J_T^{\mu\nu} ,
\end{eqnarray}
we get formally (up to polynomials in $q$) the following expressions for the matrix elements:
\begin{eqnarray}
&& \hspace{-11mm}   \langle A | \Theta^{\mu\nu} | B \rangle =
\sum_S \frac{f_S}{3} \frac{g^{\mu\nu} q^2-q^\mu q^\nu}{m_S^2-q^2-i\epsilon} \langle A | J_S|B \rangle  \nonumber \\
&& + \sum_T  f_T \frac{m_T^2} {m_T^2-q^2-i\epsilon} \langle A | \sum_\lambda \epsilon^{\mu\nu}_\lambda
\epsilon^\lambda_{\alpha \beta}J_T^{\alpha \beta}|B \rangle.
\end{eqnarray}

From a field theory point of view, the above meson dominance formula should
not be taken literally, as it does not incorporate the notion of
subtractions or the pQCD high momentum behavior. Besides, it is well
known that higher spin fields have problems, particularly due to the
role played by the off-shell behavior of propagators. The
simplest way to avoid these issues is to use, in the spirit of
dispersion relations, the meson dominance for the absorptive parts, where by
construction the mesons are on the mass shell. Then, the information from
pQCD is used to apply a minimal needed number of subtractions as the short distance constraints.

The absorptive
part of the form factor in the time-like region, $q^2 \to s+ i \epsilon$,  where
we have the process $g \to R \to A \bar B$, reads
\begin{eqnarray}
\frac1{\pi} {\rm Im}  \langle A \bar B  | \Theta^{\mu \nu} | 0 \rangle = 
 \sum_R \langle A \bar B | R \rangle  \langle R | \Theta^{\mu \nu} | 0 \rangle \delta(m_R^2-s).
\end{eqnarray}
With this form, one can reconstruct the dispersive part from the dispersion relation with suitable subtraction constants.

The vacuum-to-hadron transition amplitudes are parametrized as
\begin{eqnarray}
  \langle S | \Theta^{\mu \nu} | 0 \rangle = \tfrac{1}{3} f_S  q^2 Q^{\mu\nu}, \;\;\; \langle T | \Theta^{\mu \nu} | 0 \rangle = f_T m_T^2 \epsilon^{\mu \nu}_\lambda, 
\end{eqnarray}
where $\epsilon^{\mu \nu}_\lambda $ is the spin-2 polarization tensor,
which is symmetric, $\epsilon^{\mu \nu}_\lambda = \epsilon^{\nu
  \mu}_\lambda$, traceless $g_{\mu \nu }\epsilon^{\mu \nu}_\lambda=0$,
and transverse $q_\mu \epsilon^{\mu \nu}_\lambda =0 $. The extra
factor of ${1}/{3}$ in the scalar case is conventional, chosen such
that $\langle S | \Theta| 0 \rangle = f_S q^2$. The tensor 
\begin{eqnarray}
Q^{\mu  \nu}\equiv g^{\mu \nu}-{q^\mu q^\nu}/{q^2}
\end{eqnarray} 
fulfills $Q^\mu_\mu=3$ and the conservation law $q^\mu Q_{\mu\nu}=0$.

The sum over the tensor polarizations is given by~\cite{Scadron:1968zz,Novozhilov:1975yt},
\begin{eqnarray}
  \sum_{\lambda}   \epsilon_\lambda^{\alpha \beta} \epsilon^{\mu\nu}_\lambda 
= \frac12 \left( Q^{\mu \alpha} Q^{\nu \beta} + Q^{\nu \alpha} Q^{\mu \beta}
  \right) - \frac13 Q^{\mu \nu} Q^{\alpha \beta}.
\end{eqnarray}
The on-shell condition $P \cdot q=0$ implies $P_\alpha Q^{\alpha,\beta}= P^\beta $, whereas
$\bar u' \slashed{q} u=0 $ implies $\gamma_\alpha Q^{\alpha,\beta}= \gamma^\beta $,
hence we obtain 
\begin{eqnarray}
&& \sum_{\lambda}   \epsilon_\lambda^{\alpha \beta} P_\alpha P_\beta \epsilon^{\mu\nu}_\lambda 
  = P^\mu P^\nu - \frac13 P^2 Q^{\mu \nu}, \nonumber \\ 
&&
\sum_{\lambda}   \epsilon_\lambda^{\alpha \beta} P^{ \{ \alpha} \gamma^{\beta \} }  \epsilon^{\mu\nu}_\lambda 
= P^{ \{\mu} \gamma^{\nu \} } - \frac13  Q^{\mu \nu}  \slashed{P}
\end{eqnarray}
(cf. the analogous tensor structure in Eq.~(\ref{eq:RamN})).

\subsection{Spectral properties and GFFs of the pion}

In pQCD, one derives the asymptotic formulas~\cite{Tong:2021ctu,Tong:2022zax,Liu:2024vkj}
  \begin{eqnarray}
    A(t)=-3 D(t)  \left( 1\! +\! {\cal O} (\alpha) \right) = -\frac{48 \pi \alpha(t) f_\pi^2} {t} \left( 1 \!+\! {\cal O} (\alpha) \right), 
  \end{eqnarray}
hence the dispersion relations and their ramifications hold similarly to the case of the vector form factor, cf. Eq.~(\ref{eq:vecas}).
In the present case,
Watson's theorem implies that at $ 4m_\pi^2 < s < 4 m_K^2 $ one has
  \begin{eqnarray}
    {\rm Im} \Theta (s) =  |\Theta(s)| \sin \delta_{00}(s)  \, , \qquad
        {\rm Im} A (s) =   |A(s)| \sin \delta_{02}(s).
  \end{eqnarray}
The {\it on-shell} couplings of the resonances to
the $\pi\pi$ continuum are denoted as
\begin{eqnarray}
  \langle S | \pi \pi \rangle &=& g_{S\pi\pi},  \\ 
   \langle T | \pi \pi \rangle &=& g_{T\pi\pi} \epsilon^{\alpha \beta}_\lambda P^\alpha P^\beta  
   = g_{T\pi\pi} \epsilon^{\alpha \beta}_\lambda p'^\alpha p^\beta . \nonumber
\end{eqnarray}
Thus, we get
\begin{eqnarray}
\frac1{\pi }{\rm Im} \langle \pi \pi | \Theta^{\mu\nu} | 0\rangle =
\sum_S \frac{g_{S\pi\pi} f_S}3 \delta(m_S^2-q^2) (g^{\mu\nu} q^2-
q^\mu q^\nu ) \nonumber \\ + \sum_{T,\lambda} \epsilon_\lambda^{\alpha \beta}
P^\alpha P^\beta \epsilon^{\mu\nu}_\lambda g_{T\pi\pi} f_T
\delta(m_T^2-q^2),
\end{eqnarray}
which naturally complies with a separate conservation for each spin channel
contribution when contracting with $q^{\mu}$. Therefore, in the narrow
resonance, large-$N_c$ motivated limit, the result is
\begin{eqnarray}
  \frac1{\pi}  {\rm Im} A (s) &=& \frac12 \sum_{T} g_{T\pi\pi} f_T  \delta( m_T^2-q^2) , \label{eq:del} \\
    \frac1{\pi}  {\rm Im} \Theta (s) 
  &=&  \sum_S g_{S\pi\pi} f_S m_S^2 \delta( m_S^2-q^2). \nonumber
  \end{eqnarray}
As expected, $A$ and $\Theta$ get contributions exclusively from the $2^{++}$ and $0^{++}$ states, respectively.

\begin{figure}
\begin{center}
  \includegraphics[angle=0,width=.49 \textwidth]{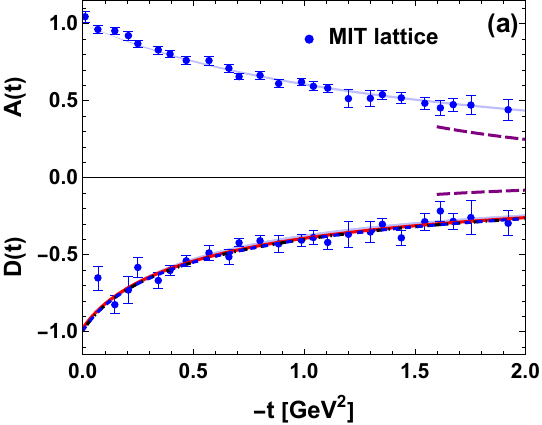} 
  \includegraphics[angle=0,width=.49 \textwidth]{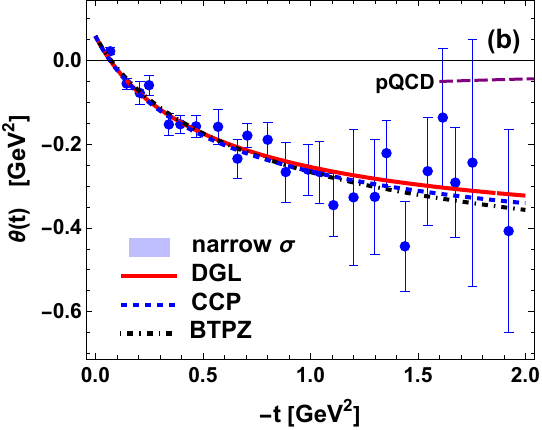} 
     \includegraphics[angle=0,width=.49 \textwidth]{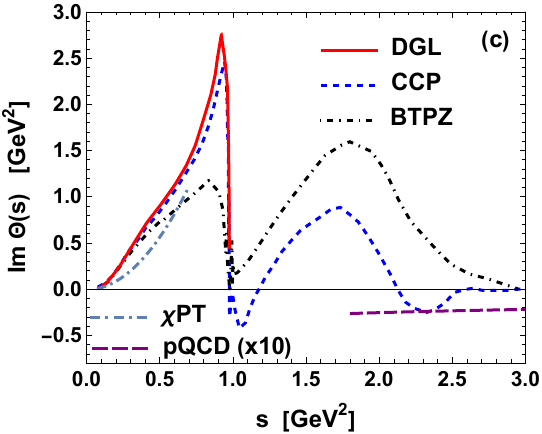} 
\end{center}
\caption{The GFFs of the pion: (a) $A(t)$ and $D(t)$, (b) $\Theta(t)$, and (c) the spectral $0^{++}$ functions. 
The legend indicates various spectral models used in the scalar sector. The long-dashed lines indicate the LO pQCD formulas, while
the blue dot-dashed line in (c) corresponds to the $\chi$PT result. \label{fig:pifit}}
\end{figure}

The minimal hadronic ansatz in a channel with good quantum numbers
corresponds to the simplest meson dominance of zero-width resonances
compatible with normalization conditions and pQCD (modulo $\alpha_S$
corrections). In our case we have~\cite{Broniowski:2024oyk}
  \begin{eqnarray}
  A (t)=\frac{m_{f_2}^{2}}{m_{f_2}^{2}-t}\, , \qquad
  \Theta (t)=2m_\pi^{2} +\frac{m_\sigma^{2} t}{m_\sigma^{2}-t}. \label{eq:ansatz0}
  \end{eqnarray}
  From here the D-term becomes
  \begin{eqnarray}
    D_\pi (0)= -1 + \frac{4 m_\pi^2}{3 m_{f_2}^2} . 
  \end{eqnarray}  
The fit to the MIT lattice data~\cite{Hackett:2023nkr}  yields $m_{f_2}^*=1.24(3)~{\rm GeV}$
and $m_\sigma^*=0.65(3)~{\rm GeV}$~\cite{Broniowski:2024oyk}. The asterisks indicate that in the
comparison/fit we have taken the lattice value of $m_\pi = 170~{\rm MeV}$. The Druck form factor is obtained from Eq.~(\ref{eq:Dpion}) 
with the earlier fitted $A(t)$ and $\Theta(t)$.
We can see from Fig.~\ref{fig:pifit} that the fit with Eq.~(\ref{eq:ansatz0}) is well within the uncertainties of the lattice determination.

To relate the form factors at {\it different} pion masses, a
mass-independent renormalization scheme is needed, such as
$\overline{\rm MS}$ in Chiral Perturbation Theory, where the so-called
gravitational low energy constants
$L_{11},L_{12},L_{13}$~\cite{Donoghue:1991qv} are needed. The analysis
with the MIT lattice data yields the following change when going from
$m_\pi=170$~MeV to the physical value of
$m_\pi=140$~MeV~\cite{Broniowski:2024oyk}:
  \begin{eqnarray} m_\sigma^*=0.65(3) \to  m_\sigma= 0.63(6), \;
     m_{f_2}^*=1.24(3) \to   m_{f_2}= 1.27(4) .
  \end{eqnarray}
The Druck term at $m_\pi=140$~MeV becomes, accordingly,
\begin{eqnarray} 
D(0) = -0.95(3).
\end{eqnarray}


\subsection{Spectral properties and GFFs of the nucleon}

From pQCD~\cite{Tong:2021ctu,Tong:2022zax} at large $Q^2=-t$, one has  
\begin{eqnarray}
  A(t) \sim + \frac{\alpha(t)^2}{(-t)^2}, \;
  J(t) \sim + \frac{\alpha(t)^2}{(-t)^2} , \;
  B(t) \sim - \frac{\alpha(t)^2}{(-t)^3} , \;
  D(t) \sim - \frac{\alpha(t)^2}{(-t)^3}. \nonumber \\ \label{eq:nasy}
\end{eqnarray}
Correspondingly, the discontinuities at large $s$ have the behavior
\begin{eqnarray}
 && {\rm Im}\,A(s) \sim + \frac{1}{s^2 L^3}, 
  {\rm Im}\,J(s) \sim +  \frac{1}{s^2 L^3},  \nonumber \\ &&
  {\rm Im}\,B(s) \sim + \frac{1}{s^3 L^3}, 
   {\rm Im}\, D(s) \sim + \frac{1}{s^3 L^3},
\end{eqnarray}
where $L=\log s/\Lambda^2_{\rm QCD}$.
This asymptotics shows that one can use the unsubtracted dispersion relations in the GFF analysis. Moreover, the 
superconvergence sum rules hold.

From Watson's theorem, in the range $ 4m_\pi^2 < s < 4 m_K^2$, one obtains
\begin{eqnarray}
&&  {\rm Im} \,\Theta (t) =  \frac{3 \sigma_\pi |f_{0,+} (t)|| \Theta_\pi (t) |}{2(4m_N^2-t)}  >0,  \nonumber \\
&&  {\rm Im } \,J (t) = \frac{3 t^2 \sigma_\pi^5}{64 \sqrt{6}} |f_{2,-} (t) || A_\pi(t) | >0, \nonumber \\
&&  {\rm Im }\,A (t) + \frac{2 t {\rm Im} J(t)}{4 m_N^2 -t} =  
  \frac{3 t^2 m_N\sigma_\pi^5}{32 \sqrt{6}} | f_{2,+} (t) || A_\pi(t) |>0, 
\end{eqnarray}  
where $\sigma_\pi= \sqrt{1-4 m_\pi^2/t}$ and $f_{l\pm}$ indicate the helicity non-flip and helicity flip amplitudes for the partial wave $l$ in the 
$\pi\pi \to N\bar N$ process. These formulas allow one to obtain the threshold behavior of GFFs from the known threshold behavior of the amplitudes 
$f_{l\pm}$~\cite{Broniowski:2025ctl}.

The {\em on-shell} couplings of the resonances to the $N \bar N$ continuum are taken as \cite{Nagels:1976mc} re-written in a suitable form 
\begin{eqnarray}
  \langle S | N \bar N \rangle &=& g_{SNN} \, , \\
  \langle T | N \bar N \rangle &=& \epsilon^{\alpha \beta}_\lambda \bar v \left[  g_{TNN}  P^{ \{ \alpha} \gamma^{\beta \} } + \frac{i\,P_{\{\mu}\sigma_{\nu\}\rho}q^\rho}{2m_N}
    f_{TNN}  \right] u
\end{eqnarray}
Thus, we get 
\begin{eqnarray}
&& \frac1{\pi }{\rm Im} \langle N  \bar N | \Theta^{\mu\nu} | 0\rangle = 
   \sum_S \frac{g_{SNN} f_S}3 \delta(m_S^2-q^2) m_S^2 Q^{\mu\nu}   \nonumber  \\ 
   & +&   \sum_{T,\lambda}   \epsilon_\lambda^{\alpha \beta}
\left[  g_{TNN}  P^{ \{ \alpha} \gamma^{\beta \} } + \frac{i\,P_{\{\mu}\sigma_{\nu\}\rho}q^\rho}{2m_N}
    f_{TNN}  \right]   
  \epsilon^{\mu\nu}_\lambda  f_T  \delta(m_T^2-q^2),
\end{eqnarray}
which naturally complies with separate conservation for each term, yielding zero when contracted
with $q^{\mu}$.  Therefore, in the narrow resonance, large-$N_c$ motivated approach we get 
\begin{eqnarray}
  \frac1{\pi}  {\rm Im} \,A (s) &=&      \sum_{T} g_{TNN} f_T  \delta( m_T^2-q^2) , \label{eq:delA} \\
    \frac1{\pi}  {\rm Im} \, B (s) &=&   \sum_{T} f_{TNN} f_T  \delta( m_T^2-q^2) , \label{eq:delB} \\
    \frac1{\pi}  {\rm Im} \,\Theta (s) 
    &=&  \sum_S g_{SNN} f_S m_S^2 \delta( m_S^2-q^2), \label{eq:delThet} 
  \end{eqnarray}
where, as expected, $A$ and $\Theta$ get contributions exclusively from the $2^{++}$ and $0^{++}$ states, respectively.  
\begin{figure}
\begin{center}
\includegraphics[angle=0,width=.49 \textwidth]{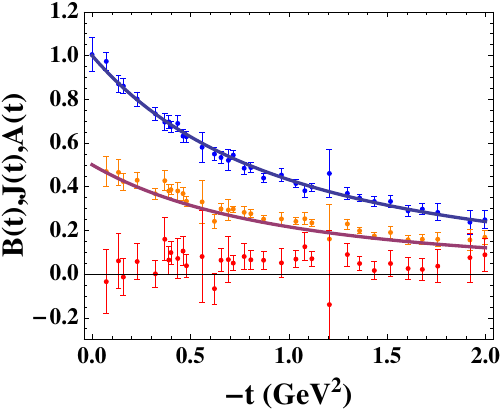}
\includegraphics[angle=0,width=.49 \textwidth]{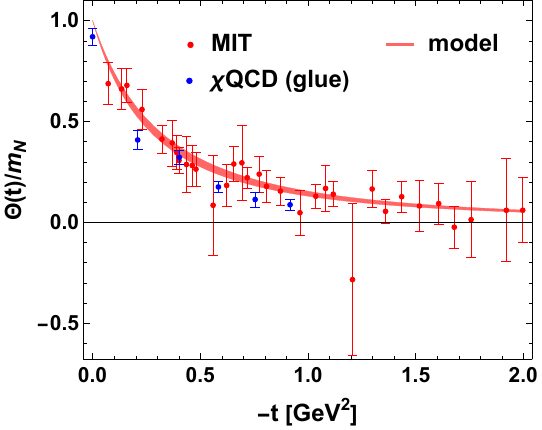}
\end{center}
\caption{The minimal hadronic ansatz for GFFs of the nucleon, compared to the MIT lattice data of~\cite{Hackett:2023rif} (points) at $m_\pi=170$~MeV. 
For additional comparison, we also display the glue contribution to the trace anomaly form factor at $m_\pi=253$~MeV from the 
$\chi$QCD Collaboration~\cite{Wang:2024lrm} (blue points in the right panel) . \label{fig:B0}}
\end{figure}
The normalization is
  \begin{eqnarray}
  A(0) = 1 \, , \qquad B(0)= 0 \, , \qquad \Theta(0)= m_N,
  \end{eqnarray}
while the high-energy behavior follows from Eq.~(\ref{eq:nasy}),
  \begin{eqnarray}
  A(t) \sim \frac{\alpha^2}{t^2} , \; B(t) \sim \frac{\alpha^2}{t^3} , \;  \Theta(t) \sim \frac{\alpha^2}{t^2}.
  \end{eqnarray}
The minimal hadronic ansatz complying to the above requirements, and
taking $B(t)=0$,\footnote{The somewhat surprising smallness of $B(t)$ (compatible with zero
  within the uncertainties of the MIT lattice data at the
  unphysical pion mass value $m_\pi=170$MeV), has been
  disclosed in~\cite{Broniowski:2025ctl}. At present, it is unclear if
  $B(t)$ becomes larger or smaller for the physical pion mass. For a
  recent attempt to explain the smallness of $B(t)$ see~\cite{Cao:2026jzm}.} yields
  \begin{eqnarray}
 && A(t) = 2 J(t) = \frac1{(1-t/m_{f_2}^2)(1-t/m_{f'_2}^2)} \, ,  \nonumber 
 \\ && B(t)=0,  \nonumber \\ && \Theta(t) =\frac{m_N}{(1-t/m_{\sigma}^2)(1-t/m_{f_0}^2)}. \label{eq:ansN}
  \end{eqnarray}
  Correspondingly the D-term becomes
\begin{eqnarray}
D_N(0)=  \frac{4 m_N^2}{3} 
   \left(-\frac{1}{m_\sigma^2}-\frac{1}{m_{f_0}^2}+\frac{1}{m_{f_2}^2}+\frac{1}{m_{f_2'}^2}\right). 
\end{eqnarray}
   We use the PDG~\cite{ParticleDataGroup:2024cfk} for the masses of $f_0 (980)$, $f_2 (1270) $, and $f_2' (1525)$ (no fit is
  carried out here) and $m_\sigma = 650(50)~{\rm MeV}$ (consistent
  with the value obtained in the pion case).  The result is shown in
  Fig.~\ref{fig:B0}. As we can see, the agreement is good, with the
  model curves falling within the error bands of the MIT data.
  Ansatz~(\ref{eq:ansN}) can be improved by relaxing the condition
  $B(t)=0$, which yields an even better agreement
  ~\cite{Broniowski:2025ctl}.

\subsection{The incompleteness problem for GFFs revisited}

Involved calculations~\cite{Cao:2024zlf} of the pion and nucleon GGFs
use the Roy equations and the Roy-Steiner equations, respectively. These are
rigorous approaches, resting on crossing, unitarity, and analyticity
below a given maximal CM energy, which typically corresponds to the $N
\bar N$ production threshold. Although the approach is theoretically quite appealing, since
the low-intermediate energy region is accurately described, the
superconvergence sum rules are broken to a large extent.
Certain phenomenological contributions mimicking infinitely many narrow resonance via the 
radial Regge trajectories may mend
the violations~\cite{RuizArriola:2026qiw}.

\section{Transverse distributions}

The physical interpretation of form factors is a subject discussed recurrently since the early works. According to our
discussion in Sect.~\ref{sec:seism}, a pictorial manner to grasp the meaning of
form factors is by looking directly at the hadron mass variation  due
to an external  gravitational field.\footnote{The same applies of course to other currents; the electroweak form factor corresponds to a variation with respect to the
electroweak fields.} Moreover,
the light-front 2D transverse viewpoint is by definition Lorentz
invariant and can be regarded as a genuinely intrinsic hadron
property.  For a detailed discussion and derivations we refer 
to~\cite{PhysRevD.15.1141,Burkardt:2000za,Diehl:2002he,Burkardt:2002hr,Miller:2010nz,Freese:2022fat,Broniowski:2024mpw,Freese:2025tqd}, where the 
transverse momentum density, energy density, and pressure are defined and discussed. Here we focus on the transverse distributions
which are naturally formulated in the light-front formulation and have
a simple partonic probabilistic interpretation.

We take the conventions $p^\pm = (p^0 \pm
p^3)/\sqrt{2}= p_\mp $, such that $x\cdot p= p^+ x^- + p^- x^+ -
p_\perp \cdot x_\perp $ and $d^4 p = dp^+ dp^- d^2 p_\perp $. Also,
for the metric tensor $\eta^{++}=\eta^{--}=0 $ and
$\eta^{+-}=\eta^{-+}=1$. Besides, we use the light-cone spinors defined
via projections
\begin{eqnarray}  
&&\Psi_\pm = {\cal P}_\pm \Psi \qquad {\cal P}_\pm =\gamma^0 \gamma^\pm = (1 \pm \gamma^0 \gamma^3)/\sqrt{2}  \nonumber \\ &&\implies 
{\cal P}_++{\cal P}_-=1 \qquad {\cal P}_{\pm}^2 = {\cal P}_\pm =
{\cal P}_\pm^\dagger, \; {\cal P}_\pm {\cal P}_\mp =0.
\end{eqnarray}

In QCD, the EM current and SEM written in LC
coordinates and in the gauge $A^+=0$ (which is ghost free) have the form
\begin{eqnarray}
J^+ &=& \Psi_+^\dagger Q \Psi_+ , \nonumber \\ 
\Theta_q^{++} &=& \frac{i}{2}  \left( \Psi^\dagger_+ \partial^+ \Psi_+ - \partial^+ \Psi_+^\dagger \Psi_+ \right), \;\;\; 
\Theta_g^{++} = (\partial^+ A^a_\perp )^2 , \nonumber \\
\Theta^{++} &=& \Theta_q^{++} + \Theta_g^{++}.
\end{eqnarray}
The field expansion for the quark field in the 
transverse coordinate space~\cite{Diehl:2002he} at $x^+=0$ is
\begin{eqnarray}
&&  q_+ (b,x^-) =  \int_0^\infty \frac{dp^+}{4\pi p^+} \sum_\lambda  \nonumber \\ && [ b_\lambda (b,p^+) u_{\lambda,+} (p^+) e^{-i p^+ x^-}  
  + d_\lambda^\dagger (b,p^+) v_{\lambda,+} (p^+) e^{i p^+ x^-} ] ,
\end{eqnarray}
with $b_\lambda^\dagger (b,p^+) $  and $d_\lambda^\dagger (b,p^+) $ denoting 
the quark and antiquark creation  operators with LC helicity $\lambda$. Then
\begin{eqnarray}
  \int d x^- q_+^+ q_+ &=& \sum_\lambda \int \frac{dp^+}{4 \pi p^+} 
  \left[ n (b,p^+) - \bar n_\lambda (b,p^+) \right], \nonumber \\
  \int d x^- q_+^+ i \partial^+ q_+ &=& \sum_\lambda \int \frac{dp^+}{4 \pi p^+} 
 \left[ p^+ n_\lambda (b,p^+) - p^+ \bar n_\lambda (b,p^+)  \right] ,
\end{eqnarray}
where  $n_\lambda (b,p^+)  = b^\dagger_\lambda (b,p^+) b_\lambda (b,p^+) $
and $ \bar n_\lambda (b,p^+)  = d^\dagger_\lambda (b,p^+) d_\lambda (b,p^+) $ denote
the particle and antiparticle number operators,  respectively. 

Thus, for  $\pi^+ = u \bar d$ taken for definiteness,
\begin{eqnarray}
  \int dx^-J^+ (b,x^-) 
\underbrace{\to}_{\pi^+} 
 \sum_\lambda \int \frac{dp^+}{4 \pi p^+} 
  \left[  \frac23  n_{u,\lambda} (b,p^+)  + \frac13  n_{\bar d,\lambda} (b,p^+) \right]. \label{eq:JJ}
\end{eqnarray}
Since $q_+^\dagger q_+ $ is positive for quarks and negative for antiquarks,
Eq.~(\ref{eq:JJ}), and consequently $F(b)$ (the Fourier transform of the 
charge form factor in the space-like momentum space, Eq.~(\ref{eq:Fb})), are positive definite.
For $\Theta_q^{++}$
one also finds positivity in an analogous way,
\begin{eqnarray}
&& \hspace{-2cm} \int dx^- \frac{i}{2}  \left(  \Psi^\dagger_+ \partial^+ \Psi_+ -\partial^+ \Psi_+^\dagger
  \Psi_+  \right) = 
  i \int dx^-   \Psi^\dagger_+ \partial^+ \Psi_+  \nonumber  \\
 \hspace{1cm}&\underbrace{\to}_{\pi^+} &
\sum_\lambda \int \frac{dp^+}{4 \pi p^+} 
  \left[  p^+ n_{u,\lambda} (b,p^+)  + p^+  n_{\bar d,\lambda} (b,p^+) \right].
\end{eqnarray}

\subsection{Wave packets on the light-front}

Consider a normalized state as a wave packet 
\begin{eqnarray}
  |\phi \rangle = \int \frac{d^2 p_\perp dp^+}{(2\pi)^3 2 p^+}
 \tilde\phi (p_\perp , p^+ ) |p_\perp , p^+ \rangle,
\end{eqnarray}
from where the scalar product is
\begin{eqnarray}
\langle \phi |\psi \rangle &=& \int \frac{d^2 p_\perp dp^+}{(2\pi)^3 2 p^+}
\tilde\phi ( p_\perp , p^+ )^* \tilde\psi (p_\perp, p^+)  \nonumber \\ &=&
\int d^2 x_\perp dx^- \phi ( x_\perp , x^- )^* \psi (x_\perp, x^-).
\end{eqnarray}
The coordinate and momentum representations are related via the Fourier transform,
\begin{eqnarray}
  \psi(x_\perp ,x^-  ) = \int \frac{d^2 p_\perp dp^+}{\sqrt{(2\pi)^3 2 p^+}} \tilde\psi (p_\perp , p^+ ) e^{i (x_\perp \cdot p_\perp - p^+ x^-)}.
\end{eqnarray}
The integration over the $x^-$ coordinate in the local operator allows one to define 
the transverse wave packet distribution in the transverse coordinate, $b=x_\perp$, as follows:
\begin{eqnarray}
  n_\psi (b) = \int dx^- |\psi(b,x^-)|^2 = \int_0^\infty  \frac{dp^+}{4 \pi p^+ } \left| \int \frac{d^2 p_\perp}{(2\pi)^2}
  e^{i b \cdot p_\perp} \tilde\psi(p_\perp , p^+) \right|^2.
\end{eqnarray}

We take the $x^+=0 $ quantization surface.
Using translational invariance, 
after some straightforward manipulations, one obtains an intuitive formula for the expectation value of the electromagnetic current $J^+$,
\begin{eqnarray}
  \langle \psi | \int dx^- J^{+} (b,x^-) | \psi \rangle &=&      \int d^2 b' n_\psi(b-b') F(b'),
  \label{eq:Jp}
\end{eqnarray}
where $F(b)$ is the Fourier transform of the charge form factor in the space-like momentum space,
\begin{eqnarray}
F(b)=\int \frac{d^2 q_\perp}{(2\pi)^2}  F (-q_\perp^2) e^{-i q_\perp \cdot b }. \label{eq:Fb}
\end{eqnarray}
For a localized wave packet $n_\psi(b) \to \delta^{(2)}(b)$ and $n^+_\psi(b) \to p^+ \delta^{(2)}(b)$, hence one has
\begin{eqnarray}
\langle \psi | \int dx^- J^+ (b,x^-) | \psi \rangle \to F(b) \,  
\end{eqnarray}
Transverse charge density is invariant under longitudinal boosts.

\subsection{Transverse distributions of GFFs}

Similarly, it is straightforward to show that $A(b)$ is the relative distribution of $P^+$ in the transverse coordinate space,
\begin{eqnarray}
&& \Theta^{++}(b) = \int \frac{d^2q_\perp}{2P^+  (2\pi)^2} e^{-i q_\perp \cdot b}  \, 2{P^+}^2 A(q_\perp^2) = 
P^+ A(b), \nonumber \\ && \int d^2b \, \Theta^{++}(b) = P^+ . \label{eq:Apos}
\end{eqnarray}
The transverse energy density is
  \begin{eqnarray}
\Theta^{+-}(b) = \int \frac{d^2q_\perp}{2P^+  (2\pi)^2} e^{-i q_\perp \cdot b}  \left [2{P^+}P^- A(q_\perp^2) + \frac{1}{2}q_\perp^2 D(q_\perp^2)\right ],
\end{eqnarray}
and does not possess definite positivity.

\subsection{Transverse densities and mechanical properties}

The form factor $D$ determines
the transverse pressure $p(b)$ and the shear forces $s(b)$ as follows~\cite{Polyakov:2002yz,Polyakov:2018zvc}:
\begin{eqnarray}
&& \Theta^{ij}(b) =  \frac{1}{2 P^+} \int \frac{d^2q_\perp}{(2\pi)^2} e^{-i q_\perp \cdot b}  \frac{1}{2}  \left [ q_\perp ^i q_\perp^j  - \delta^{ij} q_\perp^2 \right ] D(q_\perp^2) 
\nonumber \\ && = \delta^{ij} p(b)+\left [ \frac{b^i b^j}{b^2}-\frac{1}{2}\delta^{ij}\right ]  s(b). \nonumber 
\end{eqnarray}
The trace of GFF is given by 
\begin{eqnarray}
 && \Theta^\mu_\mu(b) = 2\Theta^{+-}(b)-\Theta^{11}(b)-\Theta^{22}(b) =  \epsilon(b) - 2 p(b) 
  \nonumber \\
&&~~~~~~~ \frac{1}{2P^+} \int \frac{d^2q_\perp}{ (2\pi)^2} e^{-i q_\perp \cdot b} 
\left [ 2(m_\pi^2+\tfrac{1}{4}q_\perp^2) A(q_\perp^2) + \frac{3}{2}q_\perp^2 D(q_\perp^2)\right ] \nonumber \\
&&~~=  \frac{1}{2P^+} \int \frac{d^2q_\perp}{ (2\pi)^2} e^{-i q_\perp \cdot b} \Theta(q_\perp^2) =  \frac{1}{2P^+} \Theta(b).
\end{eqnarray}
We note that $\int_0^\infty 2\pi b \, db \, p(b)=0$, as expected from classical mechanical stability. Also~\cite{Polyakov:2002yz,Polyakov:2018zvc}, 
\begin{eqnarray}
D(0)=2m_N\int_0^\infty 2\pi b \, db \, b^2 p(b).
\end{eqnarray}

\begin{figure}
\begin{center}
   \includegraphics[angle=0,width=.6 \textwidth]{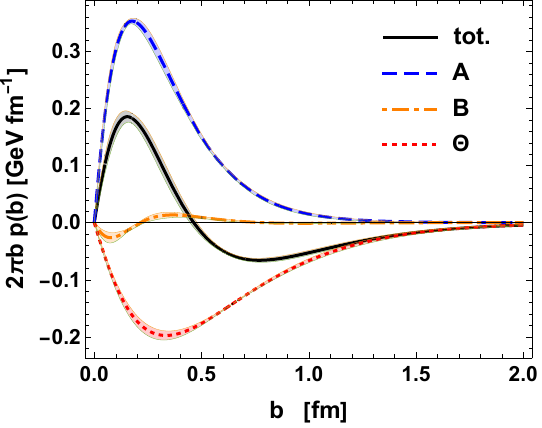}
\end{center}
\caption{Anatomy of the transverse pressure (multiplied with $2\pi b$) 
from the decomposition of Eq.~(\ref{eq:pdec}).  The attractive part comes from $\Theta$, the repulsive part from $A$,
while the contribution of $B$ is small and changes the sign. \label{fig:pdec}}
\end{figure}

Interestingly, one can decompose the pressure as follows~\cite{Broniowski:2024mpw,Broniowski:2025ctl}:
\begin{eqnarray}
p(b)=\frac{m_N}{6}A(b)+ \frac{1}{24 m_N}\nabla_b^2B(b) - \frac{1}{6}\Theta(b), \label{eq:pdec}
\end{eqnarray}
which is displayed in Fig.~\ref{fig:pdec}.
The contribution of $A$ is positive according to the general argument of Eq.~(\ref{eq:Apos}), thus repulsive and short-range, reflecting the large mass of $f_2$.
The term from $B$ is small and with no definite sign (here we use the 
meson dominance model parametrization from~\cite{Broniowski:2025ctl}, where $B$ is small but nonzero). Importantly, 
the contribution of $\Theta$ is attractive and long range, reflecting the smallness of the $\sigma$ mass.
Therefore, the meson dominance offers naturally a simple picture with a $2^{++}$ repulsion in the core and a $0^{++}$ attraction in the tail, reflecting the 
hierarchy of masses.
Qualitatively similar results were obtained in~\cite{Ji:2025gsq,Fujii:2025aip,Fujii:2025pkv}, where the $0^{++}$ component office SEM is associated with the 
gluon contribution to the trace anomaly.

\subsection{Radii}

The transverse radii are defined as
\begin{eqnarray}
\langle b^2 \rangle_F = \frac{\int_0^\infty 2\pi b \, b^2 F(b)}{\int_0^\infty 2\pi b \,  F(b)}=\frac{4}{F(0)} \left . \frac{dF(t)}{dt} \right |_{t=0}.
\end{eqnarray}
In the model parametrization of~\cite{Broniowski:2025ctl},
\begin{eqnarray}
\langle b^2 \rangle_A=4 \left ( -c_A+\frac{1}{m_{f_2}^2} +\frac{1}{m_{f'_2}^2}+\frac{1}{m_{f''_2}^2}+\frac{1}{m_{f'''_2}^2} \right ) =  {[0.34(1)~{\rm fm}]^2} \nonumber
\end{eqnarray}
$c_A$ approximately cancels the contribution ${1}/{m_{f''_2}^2}+{1}/{m_{f'''_2}^2}$
\begin{eqnarray}
\langle b^2 \rangle_\Theta=4 \left (\frac{1}{m_{\sigma}^2} +\frac{1}{m_{f_0}^2} \right ) = {[0.60(3)~{\rm fm}]^2} \nonumber
\end{eqnarray}
\begin{eqnarray}
\langle b^2 \rangle_{\rm mech}= \frac{\int_0^\infty 2\pi b \, b^2 [p(b)+\tfrac{1}{2}s(b)]}{\int_0^\infty 2\pi b [p(b)+\tfrac{1}{2}s(b)]} 
= \frac{4 D(0)}{\int_0^\infty d(-t) D(t)}=  {[0.48(3)~{\rm fm}]^2} \nonumber
\end{eqnarray}

Hierarchy of the radii reflects the meson mass pattern:
\begin{eqnarray}
&& \langle b^2 \rangle_A^{1/2}<\langle b^2 \rangle_{\rm mech}^{1/2} < \langle b^2 \rangle_\Theta^{1/2}, \\
&&  0.34(1) < 0.48(3) < 0.60(3) [{\rm fm}] . \nonumber 
\end{eqnarray}

For the case of the nucleon one can use the Abel transform~\cite{Panteleeva:2021iip,Freese:2021mzg}
to obtain the relation of the transverse (2D) and the radial (3D) distributions. 
Note that for the pion the relation does not hold, as the 3D distributions do not reflect the intrinsic structure.
The corresponding 3D radii obey the hierarchy~\cite{Broniowski:2025ctl}
\begin{eqnarray}
&& \langle r^2 \rangle^{1/2}_A < \langle r^2 \rangle^{1/2}_J < \langle r^2 \rangle^{1/2}_E < \langle r^2 \rangle^{1/2}_{\rm mech}  < \langle r^2 \rangle^{1/2}_\Theta \nonumber \\
&& 0.51(1) < 0.57(3) < 0.67(2) < 0.72(5) < 0.90(4) [{\rm fm}] \nonumber
\end{eqnarray}
As a benchmark, the charge radius of the proton is  $0.84$~fm, while its magnetic radius is $0.85$~fm.

\section{Summary}

In these lectures we have attempted to broadly discuss some general
features of the stress-energy-momentum tensor and its matrix elements,
starting from classical mechanics and classical field theory concepts,
and ending up with QCD and the meson dominance explanation of the
recent lattice data for the hadronic gravitational form factors.  We
hope to have convinced the reader of a very basic nature and a
phenomenological importance of these issues.

The principal points concerning the results for the gravitational form factors of the pion and nucleon are following:
\begin{enumerate}
\item The gravitational form factors provide insight into the matter distribution inside hadrons, in particular  the mass and forces.
They are related to the generalized parton distributions as low momentum transfers, which makes them accessible experimentally. 
\item The MIT lattice benchmark data provide high accuracy gravitational form factors for the pion and nucleon directly 
in the ``intermediate'' space-like region up to $Q^2=2~{\rm GeV}^2$.
These data are fully compatible with the meson dominance approach.
 \item It is important to carry out the form factor analysis in {good spin channels}.
 \item The matter extension (radius) is large due to the small value of the $\sigma$ meson mass, $m_\sigma=0.64(4)$~GeV.
 Precise modeling involves  a {broad $\sigma$} described with an appropriate spectral function, but the description of the space-like data is largely 
 insensitive to the spectral details. 
\item The form factor  $D(t)$ (the Druck term) is a combination of the good spin form factors, $0^ {++}$ and $2^ {++}$, with the meson dominance 
applied to the MIT lattice data yielding
   \begin{eqnarray}D_\pi(0) = -0.95(3), \, \qquad D_N(0)=-3.0(4) .
   \end{eqnarray}
   \item The gravitational transverse distributions are intrinsic properties of hadrons. 
   The meson dominance provides an efficient description of the transverse distributions at not too small transverse radii,  $b \gtrsim 0.1$~fm. 
 \end{enumerate}

\begin{table}[th]
\caption{Summary of basic properties of GFFs of the pion. \label{tab:sumpi}}
\begin{center}
{\small
\begin{tabular}{llllll}
quantity &  low limit & &intermediate range   & high limit & \\ \hline
${\rm Im}\, A(s)$  & $+$ & $2\pi$ & changes sign   & $-$ & pQCD\\
${\rm Im}\, D(s)$  & $-$ &  & changes sign   & $+$ & \\
${\rm Im}\, \Theta(s)$  & $+$ & & changes sign   & $-$ & \\ \hline
$A(-Q^2)$  &  $1$ & sym. &  & $+$ & pQCD\\
$D(-Q^2)$  &  $-1+{\cal O}(m_\pi^2)$ & &  & $-$ & \\
$\Theta(-Q^2)$  & $2m_\pi^2$ & & changes sign  & $-$ &\\ \hline
$A(b)$  & $+\infty$  & pQCD &  positive definite & $+$ & $2\pi$\\
$\Theta(b)$  & $-\infty$  & & changes sign  & $+$  & \\
$p(b)$  & $+\infty$  & & changes sign  & $-$ &
\end{tabular}
}
\end{center}
\end{table}

\begin{table}[tb]
\caption{Summary of basic properties of GFFs of the nucleon.  \label{tab:sumN}}
\begin{center}
{\small
\begin{tabular}{llllll}
quantity &  low limit & & intermediate range   & high limit &\\ \hline
${\rm Im}\, A(s)$  & $+$ & $2\pi$ & changes sign   & $+$ & pQCD\\
${\rm Im}\, J(s)$  & $+$ & & changes sign   & $+$ & \\
${\rm Im}\, B(s)$  & $+$ & & changes sign   & $+$ &\\
${\rm Im}\, D(s)$  & $-$ & & changes sign   & $+$ &\\
${\rm Im}\, \Theta(s)$  & $+$  & & changes sign   & $-$ &\\ \hline
$A(-Q^2)$  & $1$ & sym. &   & $+$ & pQCD \\
$J(-Q^2)$  & $\frac{1}{2}$ & &   & $+$ &\\
$B(-Q^2)$  & $0$ & &   & $-$ &\\
$D(-Q^2)$  &  & &   & $-$ &\\
$\Theta(-Q^2)$  & $m_N$ & & changes sign  & $-$ &\\ \hline
$A(b)$  & +  & &  positive definite & $+$ & $2\pi$ \\
$\Theta(b)$  &   & &  & $+$ & \\
$p(b)$  &   & & changes sign  & $-$ &
\end{tabular}
}
\end{center}
\end{table}

The main general features of GFFs of the pion and nucleon are collected in Tables~\ref{tab:sumpi} and \ref{tab:sumN},
containing the properties of the spectral densities, form factors for the space-like momenta, and the transverse densities.
The signs of the low- and high values of the arguments are indicated with $\pm$. 
For the case of spectral densities, ``low" means the behavior near the $2\pi$ production threshold, while ``high" denotes the asymptotic limit. 
For the remaining cases, ``low" means the zero argument. Labels pQCD, $2\pi$, and sym. indicate the reason 
for the listed behavior: perturbative QCD, the two-pion threshold, and the symmetry (the Ward-Takahashi identity), respectively.

We are grateful to the authors of Refs.~\cite{Hackett:2023rif}
and~\cite{Wang:2024lrm} for providing us the data used in the
figures. We also thank Pablo Sanchez-Puertas for many discussions. ERA
was supported by Spanish MINECO and European FEDER funds grant and
Project No.  PID2023-147072NB-I00 funded by
MCIN/AEI/10.13039/501100011033, and by the Junta de Andaluc\'\i a
grant FQM-225.


\begin{thebibliography}{100}

\bibitem{Chambers:1956zz}
E.E. Chambers and R. Hofstadter,
\newblock Phys. Rev. 103 (1956) 1454.

\bibitem{Hughes:1965zza}
E.B. Hughes et~al.,
\newblock Phys. Rev. 139 (1965) B458.

\bibitem{Feynman:1963ax}
R.P. Feynman,
\newblock Acta Phys. Polon. 24 (1963) 697.

\bibitem{DeWitt:1967uc}
B.S. DeWitt,
\newblock Phys. Rev. 162 (1967) 1239.

\bibitem{Donoghue:1994dn}
J.F. Donoghue,
\newblock Phys. Rev. D 50 (1994) 3874, gr-qc/9405057.

\bibitem{Buoninfante:2024yth}
L. Buoninfante et~al.,
\newblock (2024), 2412.08696.

\bibitem{Kobzarev:1962wt}
I.Y. Kobzarev and L.B. Okun,
\newblock Zh. Eksp. Teor. Fiz. 43 (1962) 1904.

\bibitem{sharp1963asymptotic}
D.H. Sharp and W.G. Wagner,
\newblock Physical Review 131 (1963) 2226.

\bibitem{Pagels:1966zza}
H. Pagels,
\newblock Phys. Rev. 144 (1966) 1250.

\bibitem{Raman:1971jg}
K. Raman,
\newblock Phys. Rev. D 4 (1971) 476.

\bibitem{Hare:1972pa}
M.G. Hare and G. Papini,
\newblock Can. J. Phys. 50 (1972) 1163.

\bibitem{Truong:1989my}
T.N. Truong and R.S. Willey,
\newblock Phys. Rev. D40 (1989) 3635.

\bibitem{Gasser:1990bv}
J. Gasser and U.G. Meissner,
\newblock Nucl. Phys. B 357 (1991) 90.

\bibitem{Donoghue:1991qv}
J.F. Donoghue and H. Leutwyler,
\newblock Z. Phys. C 52 (1991) 343.

\bibitem{Ji:1994av}
X.D. Ji,
\newblock Phys. Rev. Lett. 74 (1995) 1071, hep-ph/9410274.

\bibitem{Ji:1996nm}
X.D. Ji,
\newblock Phys. Rev. D 55 (1997) 7114, hep-ph/9609381.

\bibitem{Polyakov:1999gs}
M.V. Polyakov and C. Weiss,
\newblock Phys. Rev. D 60 (1999) 114017, hep-ph/9902451.

\bibitem{Polyakov:2002yz}
M.V. Polyakov,
\newblock Phys. Lett. B 555 (2003) 57, hep-ph/0210165.

\bibitem{Polyakov:2018zvc}
M.V. Polyakov and P. Schweitzer,
\newblock Int. J. Mod. Phys. A 33 (2018) 1830025, 1805.06596.

\bibitem{Hackett:2023rif}
D.C. Hackett, D.A. Pefkou and P.E. Shanahan,
\newblock (2023), 2310.08484.

\bibitem{Hackett:2023nkr}
D.C. Hackett et~al.,
\newblock Phys. Rev. D 108 (2023) 114504, 2307.11707.

\bibitem{Brommel:2007zz}
D. Brommel,
\newblock {Pion Structure from the Lattice},
\newblock PhD thesis, Regensburg U., 2007.

\bibitem{QCDSF:2007ifr}
QCDSF, UKQCD, D. Br\"ommel et~al.,
\newblock Phys. Rev. Lett. 101 (2008) 122001, 0708.2249.

\bibitem{Delmar:2024vxn}
J. Delmar et~al.,
\newblock {40th International Symposium on Lattice Field Theory}, 2024,
  2401.04080.

\bibitem{Shanahan:2018pib}
P.E. Shanahan and W. Detmold,
\newblock Phys. Rev. D 99 (2019) 014511, 1810.04626.

\bibitem{Wang:2024lrm}
\ensuremath{\chi}QCD, B. Wang et~al.,
\newblock Phys. Rev. D 109 (2024) 094504, 2401.05496.

\bibitem{Belle:2015oin}
Belle, M. Masuda et~al.,
\newblock Phys. Rev. D 93 (2016) 032003, 1508.06757.

\bibitem{Kumano:2017lhr}
S. Kumano, Q.T. Song and O.V. Teryaev,
\newblock Phys. Rev. D 97 (2018) 014020, 1711.08088.

\bibitem{CLAS:2015uuo}
CLAS, H.S. Jo et~al.,
\newblock Phys. Rev. Lett. 115 (2015) 212003, 1504.02009.

\bibitem{Burkert:2018bqq}
V.D. Burkert, L. Elouadrhiri and F.X. Girod,
\newblock Nature 557 (2018) 396.

\bibitem{GlueX:2019mkq}
GlueX, A. Ali et~al.,
\newblock Phys. Rev. Lett. 123 (2019) 072001, 1905.10811.

\bibitem{Wang:2022vhr}
X.Y. Wang, F. Zeng and Q. Wang,
\newblock Phys. Rev. D 105 (2022) 096033, 2204.07294.

\bibitem{Guo:2025jiz}
Y. Guo, F. Yuan and W. Zhao,
\newblock (2025), 2501.10532.

\bibitem{Kharzeev:2021qkd}
D.E. Kharzeev,
\newblock Phys. Rev. D 104 (2021) 054015, 2102.00110.

\bibitem{Goharipour:2025lep}
MMGPDs, M. Goharipour et~al.,
\newblock (2025), 2501.16257.

\bibitem{Song:2025zwl}
Q.T. Song, O.V. Teryaev and S. Yoshida,
\newblock Phys. Lett. B 868 (2025) 139797, 2503.11316.

\bibitem{Han:2025mvq}
J. Han, B. Pire and Q.T. Song,
\newblock Phys. Rev. D 112 (2025) 014048, 2506.09854.

\bibitem{Han:2025eao}
J. Han, B. Pire and Q.T. Song,
\newblock Phys. Rev. D 113 (2026) 014027, 2511.05970.

\bibitem{Alharazin:2026wfh}
H. Alharazin and J.Y. Panteleeva,
\newblock (2026), 2602.19267.

\bibitem{Hatta:2025ryj}
Y. Hatta and J. Schoenleber,
\newblock Phys. Rev. Lett. 134 (2025) 251901, 2502.12061.

\bibitem{Andrianov:1998fr}
Andrianov, A. A. and Andrianov, V. A. and Yudichev, V. L.
\newblock J. Math. Sci. 8 (1998) 142.

\bibitem{Megias:2004u}
Megias, E. and Ruiz Arriola, E. and Salcedo, L. L. and Broniowski, W.
\newblock Phys. Rev. D 70 (2004) 034031.

\bibitem{Megias:2005fj}
Megias, E. and Ruiz Arriola, E. and Salcedo, L. L.
\newblock Phys. Rev. D 72 (2005) 014001. 

\bibitem{Broniowski:2007si}
W. Broniowski, E. Ruiz~Arriola and K. Golec-Biernat,
\newblock Phys. Rev. D 77 (2008) 034023, 0712.1012.

\bibitem{Broniowski:2008hx}
W. Broniowski and E. Ruiz~Arriola,
\newblock Phys. Rev. D 78 (2008) 094011, 0809.1744.

\bibitem{Frederico:2009fk}
T. Frederico et~al.,
\newblock Phys. Rev. D 80 (2009) 054021, 0907.5566.

\bibitem{Masjuan:2012sk}
P. Masjuan, E. Ruiz~Arriola and W. Broniowski,
\newblock Phys. Rev. D 87 (2013) 014005, 1210.0760.

\bibitem{Fanelli:2016aqc}
C. Fanelli et~al.,
\newblock Eur. Phys. J. C 76 (2016) 253, 1603.04598.

\bibitem{Freese:2019bhb}
A. Freese and I.C. Clo\"et,
\newblock Phys. Rev. C 100 (2019) 015201, 1903.09222,
\newblock [Erratum: Phys.Rev.C 105, 059901 (2022)].

\bibitem{Krutov:2020ewr}
A.F. Krutov and V.E. Troitsky,
\newblock Phys. Rev. D 103 (2021) 014029, 2010.11640.

\bibitem{Xing:2022mvk}
Z. Xing, M. Ding and L. Chang,
\newblock Phys. Rev. D 107 (2023) L031502, 2211.06635.

\bibitem{Xu:2023izo}
Y.Z. Xu et~al.,
\newblock Eur. Phys. J. C 84 (2024) 191, 2311.14832.

\bibitem{Li:2023izn}
Y. Li and J.P. Vary,
\newblock Phys. Rev. D 109 (2024) L051501, 2312.02543.

\bibitem{Liu:2024jno}
W.Y. Liu et~al.,
\newblock Phys. Rev. D 110 (2024) 054021, 2405.14026.

\bibitem{Liu:2024vkj}
W.Y. Liu, E. Shuryak and I. Zahed,
\newblock Phys. Rev. D 110 (2024) 054022, 2405.16269.

\bibitem{Wang:2024sqg}
X. Wang et~al.,
\newblock (2024), 2406.09644.

\bibitem{Sultan:2024hep}
M.A. Sultan et~al.,
\newblock Phys. Rev. D 110 (2024) 054034, 2407.10437.

\bibitem{Fujii:2024rqd}
D. Fujii, A. Iwanaka and M. Tanaka,
\newblock (2024), 2407.21113.

\bibitem{Krutov:2024adh}
A.F. Krutov and V.E. Troitsky,
\newblock Phys. Rev. D 111 (2025) 034034, 2410.17570.

\bibitem{Choi:2025rto}
Y. Choi, H.D. Son and H.M. Choi,
\newblock Phys. Rev. D 112 (2025) 014043, 2504.14997.

\bibitem{Puhan:2025kzz}
S. Puhan et~al.,
\newblock (2025), 2504.14982.

\bibitem{Goeke:2001tz}
K. Goeke, M.V. Polyakov and M. Vanderhaeghen,
\newblock Prog. Part. Nucl. Phys. 47 (2001) 401, hep-ph/0106012.

\bibitem{Belitsky:2002jp}
A.V. Belitsky and X. Ji,
\newblock Phys. Lett. B 538 (2002) 289, hep-ph/0203276.

\bibitem{Ando:2006sk}
S.i. Ando, J.W. Chen and C.W. Kao,
\newblock Phys. Rev. D 74 (2006) 094013, hep-ph/0602200.

\bibitem{Diehl:2006ya}
M. Diehl, A. Manashov and A. Schafer,
\newblock Eur. Phys. J. A 29 (2006) 315, hep-ph/0608113,
\newblock [Erratum: Eur.Phys.J.A 56, 220 (2020)].

\bibitem{Moiseeva:2012zi}
A.M. Moiseeva and A.A. Vladimirov,
\newblock Eur. Phys. J. A 49 (2013) 23, 1208.1714.

\bibitem{Dorati:2007bk}
M. Dorati, T.A. Gail and T.R. Hemmert,
\newblock Nucl. Phys. A 798 (2008) 96, nucl-th/0703073.

\bibitem{Alharazin:2020yjv}
H. Alharazin et~al.,
\newblock Phys. Rev. D 102 (2020) 076023, 2006.05890.

\bibitem{Cebulla:2007ei}
C. Cebulla et~al.,
\newblock Nucl. Phys. A 794 (2007) 87, hep-ph/0703025.

\bibitem{Tanaka:2025pny}
M. Tanaka, D. Fujii and M. Kawaguchi,
\newblock Phys. Rev. D 112 (2025) 054048, 2507.21220.

\bibitem{Goeke:2007fp}
K. Goeke et~al.,
\newblock Phys. Rev. D 75 (2007) 094021, hep-ph/0702030.

\bibitem{Neubelt:2019sou}
M.J. Neubelt et~al.,
\newblock Phys. Rev. D 101 (2020) 034013, 1911.08906.

\bibitem{Abidin:2009hr}
Z. Abidin and C.E. Carlson,
\newblock Phys. Rev. D 79 (2009) 115003, 0903.4818.

\bibitem{Mamo:2022eui}
K.A. Mamo and I. Zahed,
\newblock Phys. Rev. D 106 (2022) 086004, 2204.08857.

\bibitem{Mondal:2015fok}
C. Mondal,
\newblock Eur. Phys. J. C 76 (2016) 74, 1511.01736.

\bibitem{Fujita:2022jus}
M. Fujita et~al.,
\newblock PTEP 2022 (2022) 093B06, 2206.06578.

\bibitem{Deng:2025fpq}
J. Deng and D. Hou,
\newblock (2025), 2502.00771.

\bibitem{Mamo:2025hur}
K.A. Mamo,
\newblock Phys. Rev. D 112 (2025) L111506, 2507.00176.

\bibitem{Mamo:2026ktr}
K.A. Mamo,
\newblock (2026), 2603.03064.

\bibitem{Nair:2024fit}
BLFQ, S. Nair et~al.,
\newblock Phys. Rev. D 110 (2024) 056027, 2403.11702.

\bibitem{Xu:2024sjt}
S. Xu et~al.,
\newblock (2024), 2408.11298.

\bibitem{Azizi:2019ytx}
K. Azizi and U. \"Ozdem,
\newblock Eur. Phys. J. C 80 (2020) 104, 1908.06143.

\bibitem{Anikin:2019kwi}
I.V. Anikin,
\newblock Phys. Rev. D 99 (2019) 094026, 1902.00094.

\bibitem{Dehghan:2025ncw}
Z. Dehghan, F. Almaksusi and K. Azizi,
\newblock (2025), 2502.16689.

\bibitem{Ji:2025gsq}
X. Ji and C. Yang,
\newblock (2025), 2503.01991.

\bibitem{Fujii:2025aip}
D. Fujii, M. Kawaguchi and M. Tanaka,
\newblock (2025), 2503.09686.

\bibitem{Kawaguchi:2025cuf}
M. Kawaguchi, M. Harada and Y.L. Ma,
\newblock Phys. Lett. B 876 (2026) 140400, 2512.23937.

\bibitem{Mejia:2025oip}
A. Mejia and P. Schweitzer,
\newblock Phys. Rev. D 113 (2026) 054016, 2511.21916.

\bibitem{Stegeman:2025sca}
R. Stegeman and R. Zwicky,
\newblock JHEP 03 (2026) 184, 2508.18537.

\bibitem{Stegeman:2025tdl}
R. Stegeman and R. Zwicky,
\newblock (2025), 2512.12315.

\bibitem{Cao:2024zlf}
X.H. Cao et~al.,
\newblock Nature Commun. 16 (2025) 6979, 2411.13398.

\bibitem{Tong:2021ctu}
X.B. Tong, J.P. Ma and F. Yuan,
\newblock Phys. Lett. B 823 (2021) 136751, 2101.02395.

\bibitem{Tong:2022zax}
X.B. Tong, J.P. Ma and F. Yuan,
\newblock JHEP 10 (2022) 046, 2203.13493.

\bibitem{Ji:2025qax}
X. Ji and C. Yang,
\newblock Nucl. Phys. B 1024 (2026) 117342, 2508.16727.

\bibitem{Broniowski:2024oyk}
W. Broniowski and E. Ruiz~Arriola,
\newblock Phys. Lett. B 859 (2024) 139138, 2405.07815.

\bibitem{RuizArriola:2024udm}
E. Ruiz~Arriola and W. Broniowski,
\newblock PoS QNP2024 (2025) 068, 2411.10354.

\bibitem{Broniowski:2024mpw}
W. Broniowski and E. Ruiz~Arriola,
\newblock Acta Phys. Polon. B 56 (2025) 3, 2412.00848.

\bibitem{Broniowski:2025ctl}
W. Broniowski and E. Ruiz~Arriola,
\newblock Phys. Rev. D 112 (2025) 054028, 2503.09297.

\bibitem{Weinberg:1972kfs}
S. Weinberg,
\newblock {Gravitation and Cosmology}: {Principles and Applications of the
  General Theory of Relativity} (John Wiley and Sons, New York, 1972).

\bibitem{sudarshan1974classical}
E.C.G. Sudarshan and N. Mukunda,
\newblock Classical dynamics: a modern perspective (World Scientific, 1974).

\bibitem{barut1980electrodynamics}
A.O. Barut,
\newblock Electrodynamics and classical theory of fields \& particles (Courier
  Corporation, 1980).

\bibitem{huang2009introduction}
K. Huang,
\newblock Introduction to statistical physics (Chapman and Hall/CRC, 2009).

\bibitem{soper2008classical}
D.E. Soper,
\newblock Classical field theory (Courier Dover Publications, 2008).

\bibitem{leutwyler1965no}
H. Leutwyler,
\newblock Il Nuovo Cimento (1955-1965) 37 (1965) 556.

\bibitem{10.1098/rstl.1884.0016}
J.H. Poynting,
\newblock Philosophical Transactions of the Royal Society of London  (1884)
  343.

\bibitem{jackson2012classical}
J.D. Jackson,
\newblock Classical electrodynamics (John Wiley \& Sons, 2012).

\bibitem{Caro:1998us}
  J. Caro, J. and L.L. Salcedo, 
\newblock Physical Review A 60 (1999) 842.

  
\bibitem{kalman1961lagrangian}
G. Kalman,
\newblock Physical Review 123 (1961) 384.

\bibitem{pitaevskii2012physical}

\newblock L.P. Pitaevskii and E.M. Lifshitz{Physical Kinetics: Volume 10}
  Vol.~10 (Butterworth-Heinemann, 2012).

\bibitem{Bjorken:1965zz}
J.D. Bjorken and S.D. Drell,
\newblock {Relativistic quantum fields. }International Series In Pure and
  Applied Physics (McGraw-Hill, New York, 1965).

\bibitem{Freedman:1974gs}
D.Z. Freedman, I.J. Muzinich and E.J. Weinberg,
\newblock Annals Phys. 87 (1974) 95.

\bibitem{Pokorski:1987ed}
S. Pokorski,
\newblock {Gauge field theories} (Cambridge University Press, 2005).

\bibitem{Fukushima:2026wwc}
K. Fukushima and T. Uji,
\newblock (2026), 2603.11704.

\bibitem{Callan:1970ze}
C.G. Callan, Jr., S.R. Coleman and R. Jackiw,
\newblock Annals Phys. 59 (1970) 42.

\bibitem{Beissner:2025nmg}
P. Bei{\ss}ner et~al.,
\newblock Eur. Phys. J. C 85 (2025) 1471, 2508.19821.

\bibitem{Deser:1967zzf}
S. Deser and D. Boulware,
\newblock J. Math. Phys. 8 (1967) 1468.

\bibitem{Suura:1973xry}
H. Suura and B.L. Young,
\newblock Phys. Rev. D 8 (1973) 4353.

\bibitem{Brout:1966oea}
R. Brout and F. Englert,
\newblock Phys. Rev. 141 (1966) 1231.

\bibitem{Bessler:1969py}
L. Bessler, T. Muta and H. Umezawa,
\newblock Phys. Rev. 180 (1969) 1604.

\bibitem{Broniowski:2022iip}
W. Broniowski, V. Shastry and E. Ruiz~Arriola,
\newblock Phys. Lett. B 840 (2023) 137872, 2211.11067.

\bibitem{Birrell:1982ix}
N.D. Birrell and P.C.W. Davies,
\newblock {Quantum Fields in Curved Space}Cambridge Monographs on Mathematical
  Physics (Cambridge Univ. Press, Cambridge, UK, 1984).

\bibitem{PhysRev.180.1604}
L. Bessler, T. Muta and H. Umezawa,
\newblock Phys. Rev. 180 (1969) 1604.

\bibitem{Won:2025dgc}
H.Y. Won and C. Lorc{\'e},
\newblock Phys. Rev. D 111 (2025) 094021, 2503.07382.

\bibitem{Ji:1995sv}
X.D. Ji,
\newblock Phys. Rev. D 52 (1995) 271, hep-ph/9502213.

\bibitem{Ji:1996ek}
X.D. Ji,
\newblock Phys. Rev. Lett. 78 (1997) 610, hep-ph/9603249.

\bibitem{Leader:2013jra}
E. Leader and C. Lorc\'e,
\newblock Phys. Rept. 541 (2014) 163, 1309.4235.

\bibitem{Lorce:2017xzd}
C. Lorc\'e,
\newblock Eur. Phys. J. C 78 (2018) 120, 1706.05853.

\bibitem{Hatta:2018sqd}
Y. Hatta, A. Rajan and K. Tanaka,
\newblock JHEP 12 (2018) 008, 1810.05116.

\bibitem{ExtendedTwistedMass:2024kjf}
Extended Twisted Mass, C. Alexandrou et~al.,
\newblock Phys. Rev. Lett. 134 (2025) 131902, 2405.08529.

\bibitem{Alexandrou:2020sml}
C. Alexandrou et~al.,
\newblock Phys. Rev. D 101 (2020) 094513, 2003.08486.

\bibitem{Novikov:1980fa}
V.A. Novikov and M.A. Shifman,
\newblock Z. Phys. C 8 (1981) 43.

\bibitem{Luscher:1984xn}
M. Luscher and P. Weisz,
\newblock Commun. Math. Phys. 98 (1985) 433,
\newblock [Erratum: Commun.Math.Phys. 98, 433 (1985)].

\bibitem{barton1965introduction}
G. Barton,
\newblock {Introduction to dispersion techniques in field theory} (W.A.
  Benjamin, New York, 1965).

\bibitem{k1969fields}
K. Nishijima,
\newblock {Fields and particles: field theory and dispersion relations} (W.A.
  Benjamin, New York, 1969).

\bibitem{Nieves:1999bx}
J. Nieves and E. Ruiz~Arriola,
\newblock Nucl. Phys. A 679 (2000) 57, hep-ph/9907469.

\bibitem{Donoghue:1996bt}
J.F. Donoghue and E.S. Na,
\newblock Phys. Rev. D 56 (1997) 7073, hep-ph/9611418.

\bibitem{Gounaris:1968mw}
G.J. Gounaris and J.J. Sakurai,
\newblock Phys. Rev. Lett. 21 (1968) 244.

\bibitem{Garcia-Martin:2011iqs}
R. Garcia-Martin et~al.,
\newblock Phys. Rev. D 83 (2011) 074004, 1102.2183.

\bibitem{tHooft:1973alw}
G. 't~Hooft,
\newblock Nucl. Phys. B 72 (1974) 461.

\bibitem{Witten:1979kh}
E. Witten,
\newblock Nucl. Phys. B 160 (1979) 57.

\bibitem{ParticleDataGroup:2024cfk}
Particle Data Group, S. Navas et~al.,
\newblock Phys. Rev. D 110 (2024) 030001.

\bibitem{RuizArriola:2011nrw}
Ruiz Arriola, Enrique and Broniowski, Wojciech. 
\newblock Proceedings of Mini-Workshop Bled 2011: Understanding Hadronic Spectra.

\bibitem{Masjuan:2017fzu}
P. Masjuan and E. Ruiz Arriola,
\newblock Phys. Rev. D 96 (2017) 054006


\bibitem{RuizArriola:2017ggc}
E. Ruiz Arriola, P. Masjuan and W. Broniowski,
\newblock Acta Phys. Polon. Supp. 10 (2017) 1079.



\bibitem{RuizArriola:2024gwb}
E. Ruiz~Arriola and P. Sanchez-Puertas,
\newblock (2024), 2403.07121.

\bibitem{BaBar:2012bdw}
BaBar, J.P. Lees et~al.,
\newblock Phys. Rev. D 86 (2012) 032013, 1205.2228.

\bibitem{sakurai1969currents}
J.J. Sakurai,
\newblock Currents and mesons (University of Chicago press, 1969).

\bibitem{Krolikowski:1967ryy}
W. Kr\'olikowski,
\newblock Phys. Lett. B 24 (1967) 305.

\bibitem{Raman:1970wq}
K. Raman,
\newblock    Phys.Rev.D 2 (1970) 1577. 

\bibitem{Raman:1971ur}
K. Raman,
\newblock Phys. Rev. D 3 (1971) 2900.

\bibitem{Toublan:1995bk}
D. Toublan,
\newblock Phys. Rev. D 53 (1996) 6602, hep-ph/9509217,
\newblock [Erratum: Phys.Rev.D 57, 4495 (1998)].

\bibitem{Ecker:2007us}
G. Ecker and C. Zauner,
\newblock Eur. Phys. J. C 52 (2007) 315, 0705.0624.

\bibitem{Scadron:1968zz}
M.D. Scadron,
\newblock Phys. Rev. 165 (1968) 1640.

\bibitem{Novozhilov:1975yt}
Y.V. Novozhilov,
\newblock {Introduction to Elementary Particle Theory}International Series of
  Monographs In Natural Philosophy (Pergamon Press, Oxford, UK, 1975).

\bibitem{Nagels:1976mc}
M.M. Nagels et~al.,
\newblock Nucl. Phys. B 109 (1976) 1.

\bibitem{Cao:2026jzm}
X. Cao et~al.,
\newblock (2026), 2601.19141.

\bibitem{RuizArriola:2026qiw}
E. Ruiz~Arriola, P. Sanchez-Puertas and W. Broniowski,
\newblock 2026, 2604.09185.

\bibitem{PhysRevD.15.1141}
D.E. Soper,
\newblock Phys. Rev. D 15 (1977) 1141.

\bibitem{Burkardt:2000za}
M. Burkardt,
\newblock Phys. Rev. D 62 (2000) 071503, hep-ph/0005108,
\newblock [Erratum: Phys.Rev.D 66, 119903 (2002)].

\bibitem{Diehl:2002he}
M. Diehl,
\newblock Eur. Phys. J. C 25 (2002) 223, hep-ph/0205208,
\newblock [Erratum: Eur.Phys.J.C 31, 277--278 (2003)].

\bibitem{Burkardt:2002hr}
M. Burkardt,
\newblock Int. J. Mod. Phys. A 18 (2003) 173, hep-ph/0207047.

\bibitem{Miller:2010nz}
G.A. Miller,
\newblock Ann. Rev. Nucl. Part. Sci. 60 (2010) 1, 1002.0355.

\bibitem{Freese:2022fat}
A. Freese and G.A. Miller,
\newblock Phys. Rev. D 108 (2023) 034008, 2210.03807.

\bibitem{Freese:2025tqd}
A. Freese,
\newblock Phys. Rev. D 112 (2025) 034037, 2505.06135.

\bibitem{Fujii:2025pkv}
D. Fujii and M. Tanaka,
\newblock Phys. Lett. B 870 (2025) 139872, 2507.23786.

\bibitem{Panteleeva:2021iip}
J.Y. Panteleeva and M.V. Polyakov,
\newblock Phys. Rev. D 104 (2021) 014008, 2102.10902.

\bibitem{Freese:2021mzg}
A. Freese and G.A. Miller,
\newblock Phys. Rev. D 105 (2022) 014003, 2108.03301.

\end{thebibliography}

\appendix

\section{Watson's theorem}
\label{app:watson}

We discuss the unitarity conditions in coupled channels when one
channel is closed. The goal is to address the $\pi\pi $ effect in the
processes $g^* , \gamma^* \to N\bar N $ above the $\pi\pi$ threshold
but below the $N \bar N$ threshold.  The unitarity of the S-matrix as a sum over
components reads
\[
SS^\dagger = 1 \implies \sum_n S_{in} S_{fn}^* = \delta_{if}.
\]
In general, we have
\[
S_{if} = \delta_{if} + 2 \pi i \delta(E_f-E_i) T_{if} ,
\]
which implies the generalized optical theorem
\[
T_{if}-T^*_{fi} = 2 \pi i \sum_n T_{in} T_{fn}^* \delta(E_n-E_i)  ,
\]
where $n$ are the open-channels.  Due to the time-reversal symmetry, the
S-matrix is symmetric , i. e.,  $S^T = S $. The channels may be open or
closed, such that the S-matrix acquires a block diagonal form where the
closed channels submatrix have a purely real $S$ matrix.  Therefore
\[
\sum_n S_{in} S_{n}^* = \delta_{if}.
\]
In our case $n = e^+ e^- , \pi^+ \pi^-, K \bar K , N \bar N $, etc. In
the case of only one channel open, say $\pi\pi \to \pi\pi $, we have $S_{11}=
e^{2 i \delta_{1}}$, with $\delta_{1}$ denoting the phase-shift. 

As a warm up,
let us consider first the case of two channels, $1=\pi\pi$ and $2=K
\bar K$, and deduce the unitarity condition on the transition $\pi\pi \to
K \bar K$ below the $K \bar K $ threshold. Then 
\begin{eqnarray}
&&S S^\dagger =  S S^* = \begin{pmatrix} S_{11} & S_{12} \\ S_{12} & S_{22} \end{pmatrix}
\begin{pmatrix} S_{11}^* & S_{12}^*  \\ S_{12}^*  & S_{22}^* \end{pmatrix}= 
\begin{pmatrix} 1 & 0 \\ 0  & 1 \end{pmatrix} \nonumber \\
&& \implies
\begin{cases} |S_{11}|^2 + |S_{12}|^2 = 1 \\
  S_{11} S_{12}^* + S_{12} S_{22}^* = 0 \\
  |S_{12}|^2 + |S_{22}|^2 = 1 
\end{cases}  \implies \frac{S_{12}}{S_{12}^*} = - \frac{S_{11}}{S_{22}^*}.
\end{eqnarray}
Using the fact that channel 2 is closed, $S_{22}= S_{22}^*$, and $S_{11}= |S_{11}| e^{2 i \delta_1}  \equiv \eta_1 e^{2 i \delta_1}$, we find that
\[
\frac{S_{12}}{S_{12}^*} = 
-\frac{\eta_1 e^{2 i \delta_{1} }}{S_{22}} \implies S_{12} = \pm i |S_{12}| e^{i \delta_{1}}.
\]
Therefore, in this case we have 
\[
S = \begin{pmatrix} \eta_1 e^{2i \delta_{1}} & + i \sqrt{1-\eta_1^2} e^{i \delta_1}
  \\  + i \sqrt{1-\eta_1^2} e^{i \delta_1}  & \eta_1 \end{pmatrix},
\]
which corresponds to the case of both channels opened,
\[
S = \begin{pmatrix} \eta_1 e^{2i \delta_{1}} & + i \sqrt{1-\eta_1^2} e^{i (\delta_1+ \delta_2)}
  \\  + i \sqrt{1-\eta_1^2} e^{i (\delta_1+ \delta_2)}  & \eta_1 e^{2 i \delta_2} \end{pmatrix},
\]
when $\delta_2 \to 0$. 

Next, we consider the case where both channels are open, but one is weakly coupled
(for instance $1=\pi\pi$ and $2=e^+e^-$), 
such that $S_{22}= 1 + \dots $ and $S_{12}=  + i F + \dots$. Then we find to the first order in
$F$ that
\begin{eqnarray}
S S^\dagger &=&  S S^* = \begin{pmatrix} S_{11} & i F  \\ i F  & 1 \end{pmatrix}
\begin{pmatrix} S_{11}^* & - i F^*  \\ - iF^*  & 1 \end{pmatrix}= 
\begin{pmatrix} 1 & 0 \\ 0  & 1 \end{pmatrix}  \\ && \implies 
\begin{cases} |S_{11}|^2 = 1 \\
  - F^* S_{11}  + F = 0  
\end{cases}  \nonumber \\ && \implies \frac{F}{F^*} = S_{11} = e^{2 i \delta_1} \implies
F = |F| e^{i \delta_1} \implies  {\rm Im} F = |F| \sin \delta_1 , \nonumber
\end{eqnarray}
which is Watson's theorem for one channel. Note that for attractive interactions $\delta_1> 0$, hence ${\rm Im} F > 0$.

\begin{figure}  
\begin{center}
  \includegraphics[angle=0,width=0.5\textwidth]{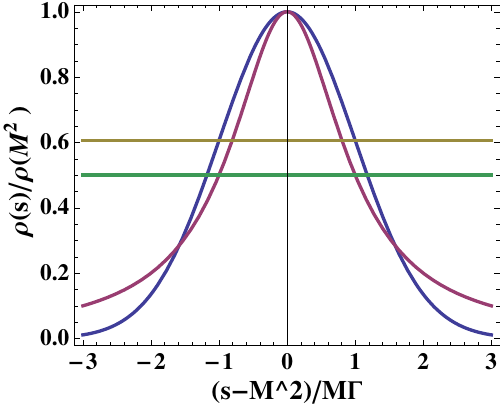}
\end{center}
\caption{Illustration of the half-width rule comparing a Gaussian and a BW distribution.}
\label{fig:hw-rule}
\end{figure}

\section{The half-width rule}
\label{app:hwr}

As we have mentioned, PDG~\cite{ParticleDataGroup:2024cfk} provides
a summary of estimates of resonance masses and widths for mesons with
given $J^{PC}$, but the more detailed information about phase-shifts
is not always available in the literature. So, what is a reasonable
numerical mass value we have to take when mapping a resonance into a
monopole form factor ?  A rather conservative estimate of the
uncertainty is given by the {\it half-width rule}. Quite generally, we
have
\[  Amplitude = Background + Resonance. \]
Realistically, one would thus have a sum of the two contributions, 
\begin{eqnarray}
  \rho(s) =\rho_B(s) + Z_R \rho_R(s) .
\end{eqnarray}
For a BW resonance profile
\begin{eqnarray}
\rho_R (s) = \frac1{\pi} \frac{M \Gamma}{(M^2-s)^2 + \Gamma^2 M^2} \, , \qquad
\int \rho (s) = 1,
\end{eqnarray}
where the normalization assumes that we integrate for simplicity over the whole axis. 
In a probabilistic interpretation we have  
\begin{eqnarray}
\frac{\rho( M_R^2 \pm \Gamma_R M_R)}{\rho(M_R^2)}=\frac12.
\end{eqnarray}
In Fig.~\ref{fig:hw-rule} we compare a Gaussian and BW shapes. As we
can see, the line shapes are very similar within the half-width rule
interval.

\end{document}